\documentclass[a4paper]{statsoc}
\usepackage{amsmath}
\usepackage{url}
\usepackage{float}
\usepackage{enumerate}
\usepackage{natbib}

\usepackage{graphics}
\usepackage{graphicx,color}
\newcommand{\erdos}{Erd\H{o}s-R\'enyi }
\usepackage{amsfonts, animate,subfigure}
\usepackage{amssymb}
\RequirePackage[colorlinks,citecolor=blue,urlcolor=blue]{hyperref}
\usepackage{todonotes}

\usepackage{indentfirst}
\usepackage{float}
\usepackage{amsfonts}
\usepackage{bm,pbox}
\usepackage{multirow} 
\usepackage{color}
\usepackage[margin=.9in]{geometry}
\usepackage{framed}
\usepackage{amssymb, amsfonts, amsbsy, latexsym, dsfont}
\usepackage{enumerate}
\newtheorem{theorem}{Theorem}

\title[Detecting Change in Temporal Networks with the DCSBM]{Modeling and Detecting Change in Temporal Networks via a Dynamic Degree Corrected Stochastic Block Model}

\author[James D. Wilson {\it et al.}]{James D. Wilson}
\address{Department of Mathematics and Statistics,
	University of San Francisco. 
	San Francisco, CA 94117}
\email{jdwilson4@usfca.edu}
\author{Nathaniel T. Stevens}
\address{Department of Mathematics and Statistics,
	University of San Francisco. 
	San Francisco, CA 94117}
\email{ntstevens@usfca.edu}
\author[James D. Wilson {\it et al.}]{William H. Woodall}
\address{Department of Statistics,
		Virginia Tech.
		Blacksburg, VA 24061}
\email{bwoodall@vt.edu}
\begin{document}
		

\begin{abstract}
\noindent In many applications it is of interest to identify anomalous behavior within a dynamic interacting system. Such anomalous interactions are reflected by structural changes in the network representation of the system. We propose and investigate the use of a dynamic version of the degree corrected stochastic block model (DCSBM) to model and monitor dynamic networks that undergo a significant structural change. We apply statistical process monitoring techniques to the estimated parameters of the DCSBM to identify significant structural changes in the network. Application of our surveillance strategy to the dynamic U.S. Senate co-voting network reveals that we are able to detect significant changes in the network that reflect both times of cohesion and times of polarization among Republican and Democratic party members. These findings provide valuable insight about the evolution of the bipartisan political system in the United States. Our analysis demonstrates that the dynamic DCSBM monitoring procedure effectively detects local and global structural changes in dynamic networks. The DCSBM approach is an example of a more general framework that combines parametric random graph models and statistical process monitoring techniques for network surveillance.

\end{abstract}



\section{Introduction}\label{sec:intro}

Time-varying, or dynamic, networks are often used to model the interactions of a group of actors through time. In many applications, it is of interest to identify anomalous behavior among the actors within a dynamic network. For example, organizers of the Arab Spring uprisings in 2012 tended to interact with one another more frequently on Facebook at the onset of the uprisings \citep{nytimes}. Similarly, central players in the ENRON scandal exchanged an increased number of emails prior to fraud investigations \citep{shetty2005discovering}. In both of these examples, anomalous activity occurred among the {\it interactions} of the actors of the system; as a result, these changes can be observed in the network describing the actors. 

The monitoring of dynamic networks for anomalous changes through time is known as \emph{network surveillance}. Network surveillance techniques have been successfully applied in a number of settings, including the detection of fraud in large online networks \citep{chau2006detecting, pandit2007netprobe, akoglu2013anomaly}, the identification of central players in terrorist groups \citep{krebs2002mapping, reid2005collecting, porter2012self}, and the detection of spammers in online social networks \citep{fire2012strangers}. As recent applications of network surveillance have grown in complexity, there has been an increased interest in developing new scalable network surveillance techniques, especially in the area of social network monitoring (see \citet{savage2014anomaly}, \citet{bindu2016mining}, and \citet{woodall2016overview} for recent reviews).
A useful area to help guide network surveillance is statistical process monitoring (SPM)\footnote[1]{{Historically the field of SPM has been referred to as statistical process control, but recently many replace the word ``control'' with ``monitoring''  \citep{woodall2016hunter}.}}. In general, statistical process monitoring provides a methodology for the real-time surveillance of any characteristic of interest. The philosophy behind SPM is that anomalous behavior in such a characteristic can be identified by distinguishing unusual variation from typical variation in an ordered sequence of observations. Stemming from applications in industrial manufacturing and public health surveillance, SPM has a rich history and many methods have been developed (see \citet{woodall1999spcoverview}, \citet{frisen2009surveillance} and \citet{woodall2014currentdirections} for reviews of methods and applications). 

In this article we propose a network surveillance framework that applies statistical process monitoring to the estimated parameters of a dynamic random graph model. We propose the use of a dynamic version of the degree corrected stochastic block model (DCSBM) from \citet{karrer2011stochastic}. The DCSBM is a probability distribution on the family of undirected graphs with discrete-valued edge weights. Importantly, the DCSBM dictates the propensity of connection between actors and captures two important aspects of social networks: heterogeneous connectivity, and community structure. As many monitoring applications involve social communications, e.g., the terrorist networks in \citet{pandit2007netprobe} and \citet{akoglu2013anomaly}, the DCSBM can be used to simulate realistic networks. 

The DCSBM is characterized by parameters for which closed-form maximum likelihood estimators (MLEs) can be readily derived. We use statistical process monitoring to identify time points at which the parameter estimates of the DCSBM change. Here, we investigate two widely-studied SPM methods for surveillance, the Shewhart control chart for individual observations and the exponentially weighted moving average (EWMA) control chart  \citep{montgomery2013quality}. We apply our surveillance strategy to the dynamic co-voting network of the U.S. Senate, which models the voting behavior of U.S. Senators from 1867 to 2015. We find that our surveillance strategy is able to identify eras of cohesion \emph{and} division among the Republican and Democrat parties, and that these changes coincide with significant political events in U.S. history. This analysis, as well as our simulation study, reveals that our network surveillance method with the DCSBM is an effective monitoring strategy for dynamic networks that undergo change.



%
%

Our proposed monitoring strategy establishes one practically useful technique among a general family of methods for surveillance. Our framework relies on two components: a parametric dynamic random graph model for modeling the features of the graph, and control charts from statistical process monitoring for the detection of changes in the parameters. Here, we consider a dynamic DCSBM random graph model and the Shewhart and EWMA control charts for surveillance. However, this same framework can be used for any parametric dynamic random graph model and any control chart of the user's choice. For example, one could investigate dynamic exponential random graph models like those described by \cite{hanneke2010discrete} and \cite{krivitsky2014separable}, or dynamic latent space models such as that introduced in \cite{sewell2015latent}. Furthermore, one could further investigate the use of other univariate SPM methods such as cumulative sum (CUSUM) control charts or control charts for attributes and perhaps multivariate SPM approaches such as Hotelling $T^2$ or multivariate EWMA control charts \citep{montgomery2013quality}. Our current proposal serves only as a first step in understanding the utility of our proposed framework. 
  
The remainder of this manuscript is organized as follows. In Section \ref{sec:surveillance_specifics} we describe the network surveillance problem in detail and discuss general approaches in the area. Section \ref{sec:related_work} describes related model-based surveillance approaches. Section \ref{sec:model} provides a description of the degree corrected stochastic block model for networks with discrete-valued edges, and how to simulate dynamic DCSBMs with structural change. Next we discuss how to estimate and monitor the DCSBM using SPM techniques in Section \ref{sec:simulation}. In Sections \ref{sec:voting_network} and \ref{sec:numerical_study} we investigate the utility of our proposed model and surveillance strategy on simulated networks and through application to the U.S. Senate co-voting network. We end with a discussion of open areas for future research in Section \ref{sec:discussion}.  

\section{The Network Surveillance Problem}\label{sec:surveillance_specifics}

Consider a collection of actors or individuals $[n] = \{1, \ldots, n\}$, whose interactions have been recorded at times $t = 1, \ldots, m$. In many applications, it is convenient to represent the interactions of $[n]$ at time $t$ by an undirected network $G_t = ([n], W_t)$. Here, the actors $[n]$ are treated as \emph{nodes} or \emph{vertices} in the graph, and $W_t = \{w_{u,v}(t): u,v \in [n]\}$ is the set of \emph{edge weights}, where $w_{u,v}(t)$ quantifies the strength of the relationship between nodes $u$ and $v$ at time $t$. A dynamic network model of the individuals $[n]$ over time $t = 1, \ldots, m$ is the ordered sequence of undirected graphs $\bm{G}(n, m) = \{G_1, \ldots, G_m\}$. The edge weight $w_{\{u,v\}}(t)$ may, for example, represent the number of communications between individuals $u$ and $v$ at time $t$ in a dynamic social network, or the number of interactions between two genes $u$ and $v$ at time $t$ in a biological network. Note that an unweighted graph, where each edge weight is binary, is a special case where edges indicate the presence or absence of a specified level of connection between nodes $u$ and $v$ at time $t$.

The goal of network surveillance is to prospectively monitor the interactions of $[n]$ so as to detect abnormal behavior among the actors. To perform surveillance, one generally first specifies a statistic ${S}_t$, or more generally a vector of statistics $\bm{S}_t$, that provides some local or global summary of the network $G_t$ based on the types of anomalies to be detected. The choice of ${S}_t$ is flexible. In the simplest case, one can choose a statistic that summarizes some topological aspect of $G_t$, such as the connectivity of each node, the clustering of nodes, or the average shortest distance between each pair of nodes \citep{priebe2005scan, marchette2012scan, neil2013scan, park2013anomaly}. In many cases, the choice of statistic is driven by the application, such as the Enron email network analysis in \citet{priebe2005scan}. Alternatively, one can model $G_t$ by a family of probability distributions governed by parameters {$\bm{\Psi}$}. In this case, one may specify $S_t$ as an estimator, or some likelihood ratio statistic, associated with {$\bm{\Psi}$}. We discuss these model-based approaches in more detail in Section \ref{sec:related_work}. 

{Once a statistic $S_t$ has been chosen, SPM is used to distinguish unusual behavior from typical behavior. In network surveillance, this corresponds to the real-time identification of unusually large or small values of ${S}_t$. The most popular technique used to determine the extremity of $S_t$ is a \emph{control chart} -- a time series plot of $S_t$ constructed with \emph{control limits} that indicate boundaries of typical behavior. An observed value of ${S}_t$ is considered anomalous if it deviates significantly from what previous observations suggest is typical. Monitoring consists of two phases, \emph{Phase I} and \emph{Phase II}, which are described below.} \\  

\noindent {\bf Phase I}: The statistic ${S}_t$ is calculated for all graphs $G_t \in \bm{G}(n,m)$. The mean $\mu$ and variance $\sigma^2$ of $S_t$ are estimated using the $m$ sampled statistics. A tolerance region $\mathcal{R}(\widehat\mu, \widehat\sigma^2)$ is constructed based on the estimated values for $\mu$ and $\sigma^2$. The upper and lower bounds of this region are referred to as upper and lower control limits, respectively. Variation within these limits defines typical behavior.\\ 

\noindent {\bf Phase II}: For each new graph $G_t$, with $t > m$, $S_t$ is calculated, and $G_t$ is deemed ``typical'' if $S_t \in \mathcal{R}(\widehat\mu, \widehat\sigma^2)$ and deemed ``anomalous'' otherwise. When an observed value of $S_t$ exceeds these limits, we say that the control chart has \emph{signalled}; this serves as an indication that a structural change has occurred.\\

\begin{figure}[ht]
	\centering
	\includegraphics[width = 0.8\textwidth]{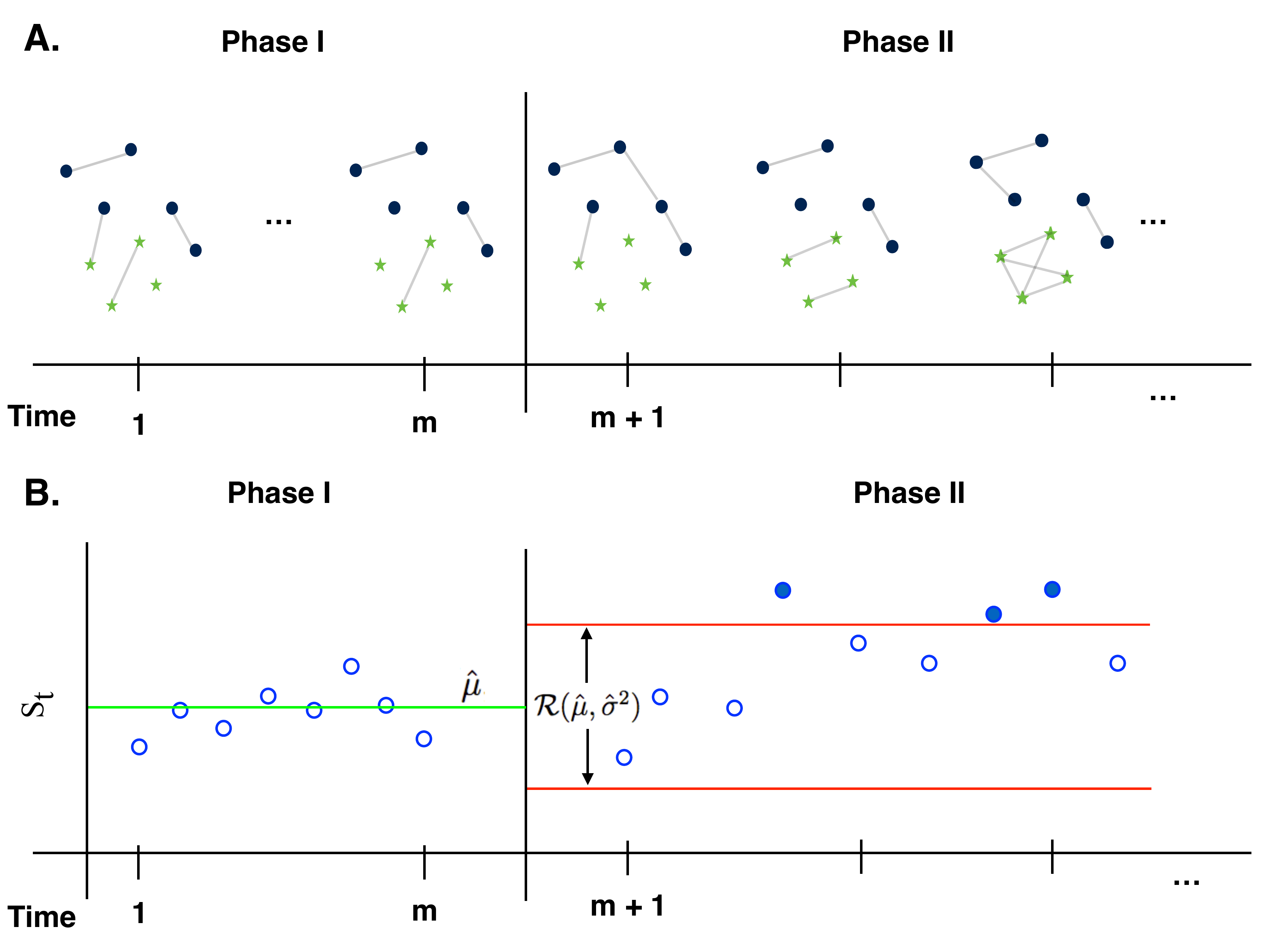}
	\caption{\label{fig:toy} {Toy example illustrating network surveillance using the statistic $S_t$, and the distinction between Phase I and Phase II}.}
\end{figure}

Data collected within Phase I serves as a baseline to establish what defines ``typical'' variation in $S_t$. Prospective monitoring begins in Phase II. For $t > m$, we formally decide whether the graph $G_t$ demonstrates anomalous behavior by comparing $S_t$ to the control limits defined in Phase I. Figure \ref{fig:toy} illustrates a toy example of this procedure.

As $\bm{G}(n,m)$ is used to determine the tolerance region $\mathcal{R}(\widehat\mu, \widehat\sigma^2)$, successful monitoring in Phase II requires that the data in Phase I provide an accurate representation of typical variation; if $\mu$ and $\sigma^2$ are not accurately estimated, then the control limits defined by $\mathcal{R}(\widehat\mu, \widehat\sigma^2)$ are unlikely to be applicable beyond the Phase I time frame. Ideally the control limits will balance the need for a control chart that is sensitive enough to detect important changes, while not signalling too frequently and creating an excessive number of false alarms. \citet{jonesfarmer2014phaseI} discuss the importance of effectively collecting and analyzing baseline data during Phase I. If the network being monitored is expected to evolve over time, then we recommend moving window approaches as opposed to a fixed Phase I sample as in \citet{zhao2016movingwindow}. 

The performance of a surveillance technique depends also on the definition of $\mathcal{R}(\widehat\mu, \widehat\sigma^2)$, which largely depends on the goal of the control chart and the type of data being plotted. Abnormal activity in Phase II networks may be brief - where as few as one or two anomalous graphs are observed, or it may persist over an extended period of time. To detect sudden large changes a standard Shewhart control chart is typically used \citep{montgomery2013quality}. However, if sensitivity to sustained small and medium-sized changes is of interest, one might consider using an exponentially weight moving average (EWMA) control chart. See \citet{saleh2015ewma} and \citet{sparks2016monitoring} for recent advances in EWMA control chart techniques.

In practice, the choice of statistic $S_t$ and type of control chart will depend on the types of network changes one wishes to detect. For instance if one seeks to detect a global change in the network (where there is an overall change in the structure, e.g. communications on average increase or decrease over the entire network) the choice of statistic and chart will be different than if one needs to detect a local change in the network (where there is a change in structure among some sub-graph of the network, e.g. communications on average increase or decrease within a particular community).

In Section \ref{sec:numerical_study} we use the DCSBM to simulate a variety of local and global network changes and we use a Shewhart control chart for individuals to detect these changes. The control limits for this control chart are defined as $\mathcal{R}(\widehat\mu, \widehat\sigma^2) = \widehat\mu \pm 3\widehat\sigma$. The performance of this surveillance technique will be quantified using the average run length (ARL); the average number of graphs until a signal indicates a change in the network. One would like the ARL to be small if a change has been simulated and large otherwise. \citet{mcculloh2011detecting} defined an average detection length metric, {and \citet{zhao2016movingwindow} define an average time-to-signal metric, which are both} equivalent to the ARL. We propose that our results be used as a performance benchmark, against which other surveillance techniques can be evaluated and compared. We also make recommendations regarding which statistics to use given the type of change one wishes to detect.

\section{Related Work}\label{sec:related_work}
The DCSBM generalizes several families of well-studied and widely-applied random graph models, such as the (non-degree corrected) stochastic block model from \citet{holland1983stochastic, snijders1997estimation, nowicki2001estimation}. The dynamic stochastic block model from \citet{xu2013dynamic}, like the DCSBM, can be used to model time-varying community structure in a network. However, the dynamic stochastic block model can only be applied to networks with binary edges, and does not address degree hetergeneity in the network. \citet{fu2009dynamic} developed a mixed membership stochastic block model, a dynamic extension to the mixed membership stochastic block model from \citet{airoldi2009mixed}, which models networks with potentially overlapping community structure. We describe the relationship of the DCSBM with several other important families of random graph models in the Appendix.

There are other model-based approaches for network surveillance that have been recently developed; we briefly describe some of them here. \citet{azarnoush2016monitoring} proposed a longitudinal logistic model that describes the (binary) occurence of an edge at time $t$ as a function of time-varying edge attributes in the sequence of networks $\bm{G}([n], T)$. This model dictates edge probabilities by the values $\bm{\beta} = (\beta_1, \ldots, \beta_p)$, where $\beta_j$ parameterizes the effect of the edge attribute $j$ on the probability of an edge. To identify anomalous behavior at time $t$, one first calculates the maximum likelihood estimates $\widehat{\bm{\beta}}_1$ and $\widehat{\bm{\beta}}_2$ for graphs $\bm{G}_1 = \{G_1, \ldots, G_{t-1}\}$ and $\bm{G}_2 = \{G_t, \ldots, G_T\}$, respectively under the longitudinal model. A likelihood ratio test is used to test the null hypothesis that $\widehat{\bm{\beta}}_1$ and $\widehat{\bm{\beta}}_2$ are equal; a rejection of the null hypothesis suggests a significant change between $\bm{G}_1$ and $\bm{G}_2$.


\citet{peel2014detecting} developed a generalized hierarchical random graph model (GHRG) to model $\bm{G}([n], T)$. To detect anomalies, the authors used the GHRG as a null model to compare observed graphs in $\bm{G}([n], T)$ via a Bayes factor. At each time $t$, Bayesian posterior inference via Markov Chain Monte Carlo is used to fit the GHRG to the graph $G_t$. Anomalies are detected using a sliding window approach on the Bayes factor that compares observed graphs to the GHRG fit for previous observations. 

In \citet{heard2010bayesian} the authors considered monitoring changes in communication volume between subgroups of targeted people over time. Their approach evaluates pairwise communication counts and determines whether these have significantly increased using a p-value. The p-value assesses the deviation of the communication rate at time $t$ and what is considered normal behavior under conjugate Bayesian models describing the discrete-valued time series of communications up to time $t$. While their focus is detecting changes on the entire network, our approach considers detecting anomalies for members of a community within a dynamic network. 

\citet{sparks2016monitoring} consider the monitoring of abrupt changes among an unknown set of actors in a dynamic network. They establish an EWMA strategy for detecting such changes, which incorporates the uncertaintly of the type and size of the subset of actors undergoing a change. In particular, they develop strategies for collaborative teams, where actors in the team each communicate more regularly; dominant leader teams, where one actor's communication greatly increases to the remainder of the team; and global outbreaks. 

The change point approach developed in \citet{barnett2016change} seeks significant changes in correlation networks, where the correlation network at time $t$ represents the correlation of some underlying multivariate stochastic process at that time. For each $t$, the Frobenius distance $F(t, t^-)$ between the correlation network at time $t$ and the average of the correlation networks from times $1, \ldots, t-1$ is calculated. The authors then generate a sample of ``null'' networks by bootstrapping a sample of $t$ networks where no change is introduced. The graph $G_t$ is said to demonstrate anomalous behavior if $F(t, t^-)$ is significantly different than the Frobenius distance under the bootstrapped sample of networks. \citet{roy2014change} considered the detection of a change point in a sequence of evolving Markov random fields. They proposed and analyzed the statistical properties of a maximum penalized pseudo-likelihood estimate, under appropriate sparsity (in the total number of edges) assumptions on the networks in $\bm{G}([n], T)$.


\section{The Degree Corrected Stochastic Block Model}\label{sec:model}

In this section we describe the degree corrected stochastic block model (DCSBM) for weighted networks. Let $G = ([n], W)$ be an undirected network that represents the interactions of actors $[n]$. The DCSBM models two important features of real networks: (i) community structure and (ii) degree heterogeneity, which we now briefly discuss.

Empirically the nodes of a network $G$ can often be divided into $k \geq 1$ disjoint vertex sets as $[n] = V_1 \cup V_2 \ldots \cup V_k$ in such a way that the number (or density) of edges within each vertex set $V_j \subseteq [n]$ is substantially greater than the number of edges between differing sets. The vertex sets are commonly referred to as \emph{communities}. In many applications, the communities of a network provide structural or functional insights about the modeled complex system. For example, recently community structure has been used to help develop hypotheses about gene interactions and antibiotic resistance \citep{parker2015network}, and about the dynamics of social interactions using cell phone data \citep{greene2010tracking}. The substantial relevance of communities in network systems has lead to a large and growing literature about community structure and the identification of statistically meaningful communities (see \citet{porter2009communities} or \citet{fortunato2010community} for reviews). 

In addition to naturally dividing into densely connected communities, actors in a network tend to have a highly variable propensity to make connections. In these situations, the \emph{degree distribution} of the nodes are variable, where the degree $d_u$ of a node $u \in [n]$ is the total number of interactions in which $u$ takes part, namely

$$ d_u = \sum_{x \in [n]} w_{u, x} .$$  

\noindent The scale-free family of networks is one common family of networks with heterogenous degrees. In scale-free networks, the degree distribution approximately follows a power law \citep{barabasi1999emergence, clauset2009power}. Scale-free networks commonly arise in economic, social, and ecological networks (e.g., \citet{kasthurirathna2015emergence} studied a recent example). The tendency of degree heterogeneity in real networks has lead to significant work in the development of fixed-degree random graph models \citep{chatterjee2011random}, and in the development of community detection methods \citep{newman2006modularity}. 

{Next in Section \ref{sec:themodel}, we first fully describe the DCSBM model for a single network, and then in Section \ref{sec:dynamic_DCSBM}, we describe how to simulate a dynamic DCSBM that undergoes a structural change. We discuss the relationship of the DCSBM with several other important random graph models in the Appendix. 
}

\subsection{The Model}\label{sec:themodel}

Let $\mathcal{G}$ represent the family of all undirected networks with $n$ nodes and $k$ disjoint communities. The DCSBM is a probability distribution $\mathbb{P}(\cdot) = \mathbb{P}(\cdot \mid \bm{\theta}, \bm{\pi}, P)$ on $\mathcal{G}$ that is characterized by (i) non-negative degree parameters $\bm{\theta} = (\theta_1, \ldots, \theta_n)$, which reflect the tendency of the nodes to connect, (ii) containment probabilities $\bm{\pi} = (\pi_1, \ldots, \pi_k)$ that satisfy $\pi_r > 0$ and $\sum_{r \in [k]} \pi_r = 1$, where $\pi_r$ specifies the probability of a node belonging to community $r$ and (iii) the $k \times k$ symmetric connectivity matrix $P = (P_{r,s})$, where entries $P_{r,s} > 0$ express the propensity of connection between nodes in communities $r$ and $s$. 


Let $\widehat{G} \in \mathcal{G}$ be a random graph with $n$ nodes and $k$ communities generated under $\mathbb{P}$. Then $\widehat{G}$ can be obtained by a simple generative procedure, which can be described as follows:

\begin{enumerate}
	\item Parameters $\bm{\theta}$, $\bm{\pi}$, and ${P}$ are pre-specified and fixed. These are chosen to control the degree variability, relative size of communities, and connection propensity between and within communities, respectively.
	\item Vertices are randomly assigned community labels $\bm{c} = (c_1, \ldots, c_n)$ according to the multinomial draws: 
	\begin{equation}\label{eq:labels} c_u \stackrel{\text{i.i.d}}{\sim} \text{Multinomial}(1, \bm{\pi}). \end{equation}
		
	\item Given $\bm{\theta}$, $\bm{c}$, and $P$, edge weights $\{w_{u,v}: u,v \in [n]\}$ are assigned according to independent Poisson draws, where
	\begin{equation}\label{eq:expectation}\mathbb{E}[w_{u,v} \mid \bm{c}, \bm{\theta}, P] = \theta_u ~ \theta_v ~ P_{c_u, c_v} \end{equation}
\end{enumerate}

The graph $\widehat{G}$ is then defined as the network with nodes $[n]$, community labels $\bm{c}$, and edge weights $\bm{w} = \{w_{u,v}: u, v \in [n]\}$ resulting from (\ref{eq:labels}) and (\ref{eq:expectation}). For an observed network with community labels $\bm{c}$ and edge weights $\bm{w}$, we define 
$$n_r = \sum_{u \in [n]} \mathbb{I}(c_u = r)$$
\noindent as the number of vertices in community $r$. Further we define
$$m_{r,s} = \sum_{u: c_u = r}\sum_{v: c_v = s} w_{u,v}$$ 
\noindent as the total weight of edges between community $r$ and $s$ (twice the weight of edges when $r = s$). It follows by combining (\ref{eq:labels}) and (\ref{eq:expectation}) that the joint distribution of the random graph $\widehat{G}$ and community labels $\bm{C}$ is described by the joint probability mass function $\mathbb{P}(\cdot, \cdot)$, where when ignoring constants, 

\begin{align}\label{eq:DCSBM}
	\mathbb{P}(\widehat{G} = G, \bm{C} = \bm{c} \mid \bm{\theta}, \bm{\pi}, P) &\propto \prod_{r \in [k]} \pi_r^{n_r} \prod_{u \in [n]} \theta_u^{d_u} \prod_{r,s \in [k]} P_{r,s}^{\frac{m_{r,s}}{2}} e^{-\frac{n_r n_s P_{r,s}}{2}}\\ &\times \prod_{u < v \in [n]} \dfrac{1}{w_{u,v}!}.\nonumber
\end{align}

The distribution $\mathbb{P}(\widehat{G} = G \mid \bm{\theta}, \bm{\pi}, P)$ is obtained by summing $\mathbb{P}(\cdot, \cdot)$ in (\ref{eq:DCSBM}) over all possible realizations of $\bm{c}$. We note that the model in (\ref{eq:DCSBM}) is not identifiable without some constraint on $\bm{\theta}$ since the likelihood is unaffected by certain opposing magnitude shifts in $\bm{\theta}$ and $P$ \citep{yan2014model}. To ensure that the model is identifiable, we require that the sum of $\theta_u$ in the same community equal the number nodes in that community, namely

\begin{equation}\label{eq:identifiability} \sum_{u:c_u = r} \theta_u = n_r, \end{equation}

\noindent for all $r = 1, \ldots, k$. For simulation, it is often of interest to specify the community labels $\bm{c}$ deterministically rather than randomly as in (\ref{eq:labels}). To distinguish these assignment strategies, we will write $\mathbb{P}(\cdot \mid \bm{\theta}, \bm{c}, P)$ to represent the probability distribution of the DCSBM when the community labels are pre-specified {\it a priori}.

We now demonstrate how to simulate dynamic graphs using the DCSBM as a starting point. 

\subsection{Simulating a Dynamic DCSBM with a Structural Change}\label{sec:dynamic_DCSBM}

We are interested in simulating an ordered sequence of graphs on the vertex set $[n]$ that demonstrate various types of significant structural change. The DCSBM $\mathbb{P}(\cdot \mid \bm{\theta}, \bm{c}, P)$ provides a flexible means to model change in a random dynamic graph.
To model a dynamic graph with a significant structural change, we generate an ordered sequence of random graphs $\widehat{\bm{G}}(n, T) = \{\widehat{G}_1, \ldots \widehat{G}_{t^*}, \ldots, \widehat{G}_T\}$ according to 

\begin{equation} \label{eq:sims} \widehat{G}_t \sim \begin{cases} \mathbb{P}(G \mid \bm{\theta}, \bm{c}, P), & t < t^* \\ 
\mathbb{P}(G \mid \bm{\theta}^*, \bm{c}^*, P^*) & t \geq t^* \end{cases}.\end{equation}

By simulating $\widehat{\bm{G}}(n,T)$ as in (\ref{eq:sims}), we introduce a structural change in the graph at time $t^*$ that persists across the remaining networks in the sequence. In this way, $\widehat{\bm{G}}_1 = \{\widehat{G}_1, \ldots, \widehat{G}_{t^*-1}\}$ are simulated as ``typical'' graphs; whereas, $\widehat{\bm{G}}_2 = \{\widehat{G}_{t^*}, \ldots, \widehat{G}_{T}\}$ are ``anomalous'' graphs. The goal of a surveillance method then is to signal as quickly as possible following the time point of change $t^*$. For network monitoring simulations, we {require} $t^* > m$ so that the change occurs after Phase I. We note that in principle one can simulate networks with multiple changes, as well as networks with changes that affect a small number of networks. 

The changes $\bm{\theta} \rightarrow \bm{\theta}^*$, $\bm{c} \rightarrow \bm{c}^*$, and $P \rightarrow P^*$ each reflect a different type of structural change in the simulated dynamic network. We first describe how to simulate $\widehat{\bm{G}}(n, T)$, and then discuss the effects of each of these three types of changes. 
To simulate a dynamic network $\widehat{\bm{G}}(n,T)$ according to (\ref{eq:sims}), one can readily use the {\bf Algorithm} outlined below. 
\begin{framed}
	{\footnotesize
\begin{center}{\bf Algorithm} \end{center}
\begin{center}{\bf Simulating a dynamic DCSBM with structural change} \end{center}
\begin{itemize}
	\item[] {\bf Given}: $\bm{c}$, $\bm{c^*}$, $P$, $P^*$, $\{\delta_r, \delta^*_r \in [0,1]\}_{r = 1, \ldots, k}$
	\vskip .15pc
	\item[] {\bf Step I}: For $t = 1, \ldots, t^*-1$
	\begin{itemize}
		\item Generate propensity parameters
		{$ \theta^{(0)}_u \stackrel{i.i.d}{\sim} U(1 - \delta_{c_u}, 1 + \delta_{c_u}) $}
		\item Scale $\theta^{(0)}_u$ values to ensure identifiability:
		$$\theta_u = \dfrac{n_{c_u} \theta^{(0)}_u}{\displaystyle\sum_{v: c_v = c_u} \theta^{(0)}_v}$$ 
		\item Generate edges of $\widehat{G}_t$ as independent Poisson draws
		$$ {w}_{u,v} \sim \text{Poisson}\left(\theta_u ~ \theta_v ~ P_{c_u, c_v}\right)$$
	\end{itemize}
	\item[] {\bf Step II}: For $t = t^*, \ldots, T$
	\begin{itemize}
		\item Repeat {\bf Step I} with updated parameters: $P \rightarrow P^*$, $\delta_r \rightarrow \delta_r^*$, and $\bm{c} \rightarrow \bm{c}^*$
	\end{itemize}
\end{itemize}
}
\end{framed}
\vskip .5pc

We choose the uniform random variable to simulate $\mathbf{\theta}$ to (i) induce stochastic variability of the degree sequence of the graphs through time and (ii) parameterize the mean and variability of the propensity of connection of the nodes within community $r$ with a single parameter $\delta_r$. In practice, any non-negative continuous or discrete random variable with finite mean and variance can be used here and depends on the application. For example, if one observes that the degree sequence is constant through time when the process is stable, then one can simulate $\mathbf{\theta}$ once and use the same values for each graph in the dynamic sequence. 
 
By altering the parameters that dictate the DCSBM from time $t^* - 1$ to $t^*$, we are able to model several types of structural change among the actors $[n]$ in $\widehat{\bm{G}}(n,T)$, including the following:

\begin{itemize}
	
	\item[(i)] {\it Change in rates of interaction}: In general, one can introduce a mean shift in interaction rate in community $r$ by specifying $P^*_{r,r} \neq P_{r,r}$. Doing so will also affect the variance of the interaction rate in the community. In particular, the mean and variance of the number of interactions in community $r$ will decrease at time $t^*$ when $P^*_{r,r} < P_{r,r}$, and increase when $P^*_{r,r} > P_{r,r}$. One can introduce a change in variance of the interaction rate in community $r$ by specifying $\delta^*_r \neq \delta_r$; in particular, this variance will increase if $\delta^*_r > \delta_r$ and decrease if $\delta^*_r < \delta_r$.
	
	
	\item[(ii)] {\it Communication outbreaks}: In network surveillance, one is often interested in identifying ``communication outbreaks'' among the members of some sub-graph $\Omega \subseteq [n]$ in the network. A communication outbreak corresponds to an increase in the average number of interactions among the members of $\Omega$. Using the DCSBM, we can model communication outbreaks among any number of communities in the network. For example, a communication outbreak among the members of community $j$ is modeled by specifying $P^*_{r,r} > P_{r,r}$ as the mean and variance of the interactions in community $r$ will increase at time $t^*$. We can model a {\it global} communication outbreak by specifying $P^*_{r,s} > P_{r,s}$ for all $r,s \in [k]$. 
 
	\item[(iii)] {\it Change in community structure}: A change in community structure of a social network can signify an important transition in the modeled system. For example, in the political voting network we consider in Section \ref{sec:voting_network}, the community structure associated with the members of the U.S. Senate significantly changes at times of extreme polarization of the Republicans and Democrats \citep{moody2013portrait}. \citet{chen2012community} describe six general types of community structure changes in a network, including growth, shrinkage, birth, death, the merging of two communities, or the splitting of a single community into two or more communities. In general, each of these types of changes can be implemented at time $t^*$ by specifying new community labels $\bm{c}^* \neq \bm{c}$. 
\end{itemize} 

Using the DCSBM, we are able to generate a dynamic random graph $\widehat{\bm{G}}([n], T)$ that reflects a structural change at time $t^*$. In this way, we can use $\widehat{\bm{G}}([n], T)$ as a ground truth on which one can assess the strengths and weaknesses of any network surveillance method.

\section{Monitoring the Dynamic DCSBM}\label{sec:simulation}

Suppose that we observe a dynamic graph sequence $\bm{G}(n, T) = \{G_1, \ldots, G_T\}$ that is generated under the dynamic DCSBM according to (\ref{eq:sims}). Our goal is to identify time points at which there is a change in the distribution that generated $\bm{G}(n, T)$. To detect such changes, we propose a surveillance strategy that proceeds in two steps. First, the dynamic DCSBM is fitted to $\bm{G}(n, T)$ using maximum likelihood estimation. Next, control charts are applied to functions of these maximum likelihood estimators to detect changes. In general, any control chart can be used to detect changes and indeed this should be further explored in future work; however, in this manuscript we consider the use of the Shewhart and EWMA control charts for individuals. We first describe estimation of the DCSBM and then our monitoring strategy. 

\subsection{Fitting the dynamic DCSBM} \label{sec:properties}

\subsubsection{Estimation of Communities} \label{sec:community_est}

The estimation of the community labels $\bm{c}$, otherwise known as \emph{community detection}, is known to be an NP hard problem; as a result one must estimate the labels using an approximate algorithm. Many detection methods have been developed for weighted and unweighted networks (see \cite{porter2009communities, fortunato2010community} for reviews). The spectral clustering algorithm \citep{von2007tutorial} is particularly well-suited for this setting due to its theoretical guarantees \citep{han2015consistent, qin2013regularized, sussman2012consistent}, which we now briefly mention. 

Let $m$ denote the number of Phase I graphs in $\bm{G}(n, T)$, and assume that $m < t^*$. Define the average Phase I graph by

$$\overline{G} = \dfrac{1}{m}\sum_{j = 1}^m G_j,$$

\noindent where the sum of two graphs $G_1 = ([n], W_1)$ and $G_2 = ([n], W_2)$ is the graph with node set $[n]$ and edge weights $W_1 + W_2$. If the probability matrix $P$ has no identical rows, then spectral clustering of the graph $\overline{G}$ will provide asymptotically consistent community label estimates $\widehat{c}$, as $m \rightarrow \infty$. This is stated formally in the next theorem.

\begin{theorem}
	Let $\bm{G}(n, T) = \{G_1, \ldots, G_T\}$ be a sequence of graphs generated under the dynamic DCSBM with binary edges given by (\ref{eq:sims}), where $1 < t^* \leq T$ is the time of structural change. That is for $t < t^*$, $G_t$ is generated under the DCSBM with community labels $\mathbf{c}$, propensity parameters $\bm{\theta}$, and probability matrix $P$. Let $m < t^*$ and define $\overline{G} = \frac{1}{m}\sum_{j = 1}^m G_j$. Let $\widehat{\mathbf{c}} = (\widehat{c}_1, \ldots, \widehat{c}_n)$ denote the community label estimates obtained from applying spectral clustering to the graph $\overline{G}$. If $P$ has no identical rows and $\bm{\theta}$ satisfies the constraint in (\ref{eq:identifiability}), then up to permutation, $\widehat{\mathbf{c}} = \mathbf{c}$, a.s. as $m \rightarrow \infty$.
\end{theorem}

Theorem 1 is an immediate consequence of the main result presented in \citet{han2015consistent}. The result of the theorem suggests that if the number of Phase I graphs is large enough, we can obtain consistent estimators for the community structure for the sequence of graphs before $t^*$. This theorem suggests that one should use as many Phase I graphs as possible, but in practice the choice of $m$ depends on the judgement of the practitioner.

For monitoring purposes, we suggest using the regularized spectral method from \citet{qin2013regularized} on the Phase I graphs in the sequence and monitoring the parameter estimates conditional on the estimated community labels for the \emph{entire} sequence of graphs. As we will see, in many cases changes in the community structure will be reflected by changes in the parameter estimates describing the DCSBM. Though we do not pursue it here, future work should investigate surveillance of community labels themselves.

\subsubsection{Maximum Likelihood Estimation of Parameters}
We now briefly summarize the maximum likelihood estimation of the DCSBM, which was derived in \cite{yan2014model}. We assume that $\bm{c}$ is fixed for all $t$ and is equal to the estimators $\widehat{\bm{c}}$ obtained from spectral clustering described above. From (\ref{eq:DCSBM}), we can show that the log likelihood of $(\bm{\theta}, \bm{\pi}, P)$ given an observed graph $G = ([n], W)$ and community labels is, when ignoring constants,

\begin{align} \label{eq:loglike} \ell(\bm{\theta}, \bm{\pi}, P \mid G, \bm{c}) &\propto \sum_{r \in [k]} n_r \log (\pi_r) + \sum_{u \in [n]} d_u \log(\theta_u)\nonumber\\
	 &+ \frac{1}{2} \sum_{r,s \in [k]} \left(m_{r,s} \log (P_{r,s}) - n_r n_s P_{r,s} \right) \end{align}
	
Taking derivatives, it is readily shown from (\ref{eq:loglike}) that the maximum likelihood estimator (MLE) for each parameter has a closed-form solution. For $u \in [n]$ and $r, s \in [k]$, the maximum likelihood estimators are given by

\begin{equation}\label{eq:MLEs}
	 \widehat{\theta}_u = \dfrac{d_u}{n_r^{-1} \displaystyle\sum_{w: c_w = c_u} d_w}, \hskip 2pc \widehat{\pi}_r = \dfrac{n_r}{n}, \hskip 2pc \widehat{P}_{r,s} = \dfrac{m_{r,s}}{n_r n_s}.
\end{equation}

\subsection{Monitoring Strategy}\label{sec:strategy}
To develop a monitoring strategy that detects local and global changes in a network, we first suppose that the community labels $\bm{c}$ are fixed throughout time. Let $k$ be the number of distinct community labels. Given $\bm{c}$, we directly monitor the MLE $\widehat{P}$, where at each time $t$ we estimate the ${k \choose 2}$ unique entries of $\widehat{P}$ for graph ${G}_t$. This statistic reflects the overall connection propensity among communities. To monitor for changes in $\bm{\theta}$, one could in principle monitor each statistic $\widehat{\theta}_u$ separately; however, this leads to an unmanageable number of control charts. Instead we monitor the sample standard deviation of the estimates $\{\widehat{\theta}_j: c_j = r\}$ at each time $t$. In particular we monitor the statistic given by

\begin{equation}\label{eq:delta} s_r = \left(\dfrac{1}{n_r-1} \sum_{u: c_u = r}(\widehat{\theta}_u - 1)^2 \right)^{1/2}, \hskip .5pc \text{$r = 1, \ldots, k$}.\end{equation}
	
Our choice in using the standard deviation is motivated by the fact that subject to (\ref{eq:identifiability}), the expectation of $\{\theta_u: c_u = r\}$ is fixed to be exactly 1. Thus, we use $s_r$ to capture the variability in overall connection within community $r$.


We note that it is possible for ${\delta}_r$ to remain fixed while the propensity parameters change.  For example, in yet to be published work \citet{yu2016propensity} define a $\theta$ for each individual within a community, and treat these propensities as fixed parameters to be modeled and monitored. Their focus is the detection of change in individual connection propensities within communities.


In summary, our surveillance plan monitors ${k \choose{2}} + k$ statistics $\{\widehat{P}_{q,r}, ~ s_q: q \leq r \in [k]\}$ through time. Even though our statistics are derived with the assumption of fixed community structure, we expect these statistics to capture some community structure changes as well, since in this scenario the mean connectivity of nodes in the network will also likely change.

\subsubsection{Shewhart Control Chart}
For each of the statistics that we estimate, we use a Shewhart and EWMA control charts to determine what values indicate a significant change. Let $S_t$ be a statistic at time $t$, and let $m$ be the number of Phase I networks. For $t > m$, the Shewhart control chart for individual outcomes signals a change in the statistic if $S_t$ lies outside of the control limits $\widehat{\mu} \pm 3 \widehat{\sigma}$, where $\widehat{\mu}$ is the sample mean of the $m$ Phase I observations, and $\widehat{\sigma}$ is the moving range estimate for the standard deviation of these $m$ observations given by

$$\widehat{\sigma} = \dfrac{\sqrt{\pi}}{2(m-1)}\sum_{j = 2}^m |S_j - S_{j-1}|.$$

\noindent Note that the constant $2/\sqrt{\pi}$ is equivalent to $d_2$, the normalization constant used in the control chart literature. 

\subsubsection{EWMA Control Chart}
Whereas the Shewhart control chart is designed to detect sudden large changes in $S_t$, the width of the $\pm 3\widehat{\sigma}$ limits results in reduced sensitivity to persistent changes that are small to medium in size. In this situation the EWMA control chart is to be preferred over the Shewhart control chart. 

Instead of plotting the observed values of $S_t$ directly, for $t > m$ the EWMA control chart is a time series plot of $Z_t$, the exponentially weighted moving average of the $S_t$, where

$$Z_t = \lambda S_t + (1-\lambda)Z_{t-1},$$

\noindent $Z_0 = \widehat{\mu}$ is a common choice for the starting value of the moving average and $\lambda$ $(0 < \lambda \leq 1)$ is a smoothing constant. Through empiral investigation \citet{crowder1989ewma} provides guidance on the choice of $\lambda$ that optimizes the performance of the EWMA control chart. \citet{montgomery2013quality} suggests that values of $\lambda$ in the interval $0.05 \leq \lambda \leq 0.25$ work well in practice with $\lambda = 0.2$ being a popular choice. The control limits of the EWMA control chart are given by

$$\widehat{\mu} \pm 3\widehat{\sigma}\sqrt{\dfrac{\lambda}{(2-\lambda)}[1-(1-\lambda)^{2t}]}.$$

\noindent Note that as $t$ increases, i.e., as the number of Phase II observations increases, these control limits approach the steady-state values given by

\begin{equation}\label{eq:EWMA}\widehat{\mu} \pm 3\widehat{\sigma}\sqrt{\dfrac{\lambda}{(2-\lambda)}}.\end{equation}

If $Z_t$ lies outside these control limits, it signals that a small and persistent change has occurred. Because the current observation $S_t$ is de-emphasized in this moving average, the EWMA control chart will not signal sudden large changes as quickly as a Shewhart control chart. Thus the nature of change one wishes to detect should dictate which control chart is used. In practice, it is sensible to simultaneously monitor $S_t$ using both approaches. We explore the utility of both the Shewhart and EWMA control charts when applied to the U.S. Senate co-voting network in Section \ref{sec:voting_network} and we use simulation to investigate the detection properties of the Shewhart control chart in Section \ref{sec:numerical_study}.

\section{Application to the U.S. Senate Voting Network}\label{sec:voting_network}
We now use the DCSBM surveillance procedure to investigate the dynamic relationship between Republican and Democrat Senators in the U.S. Congress. We analyzed the co-voting network of the U.S. Senate from 1867 (Congress 40) to 2015 (Congress 113). This network was first analyzed in \citet{moody2013portrait} and has been since investigated in \citet{roy2014change}. In \citet{moody2013portrait} the modularity, or extent of divisiveness, of the network was calculated over time, and it was found that generally Republicans and Democrats have become more polarized over time. The dynamic DCSBM framework provides a means to formally model this network and test for changes in the community structure and voting patterns among party members.  

\subsection{Description of Data}

\begin{figure}[ht]
	\centering
	\includegraphics[width = \textwidth, trim = 0cm 0cm 0cm 2cm, clip = TRUE]{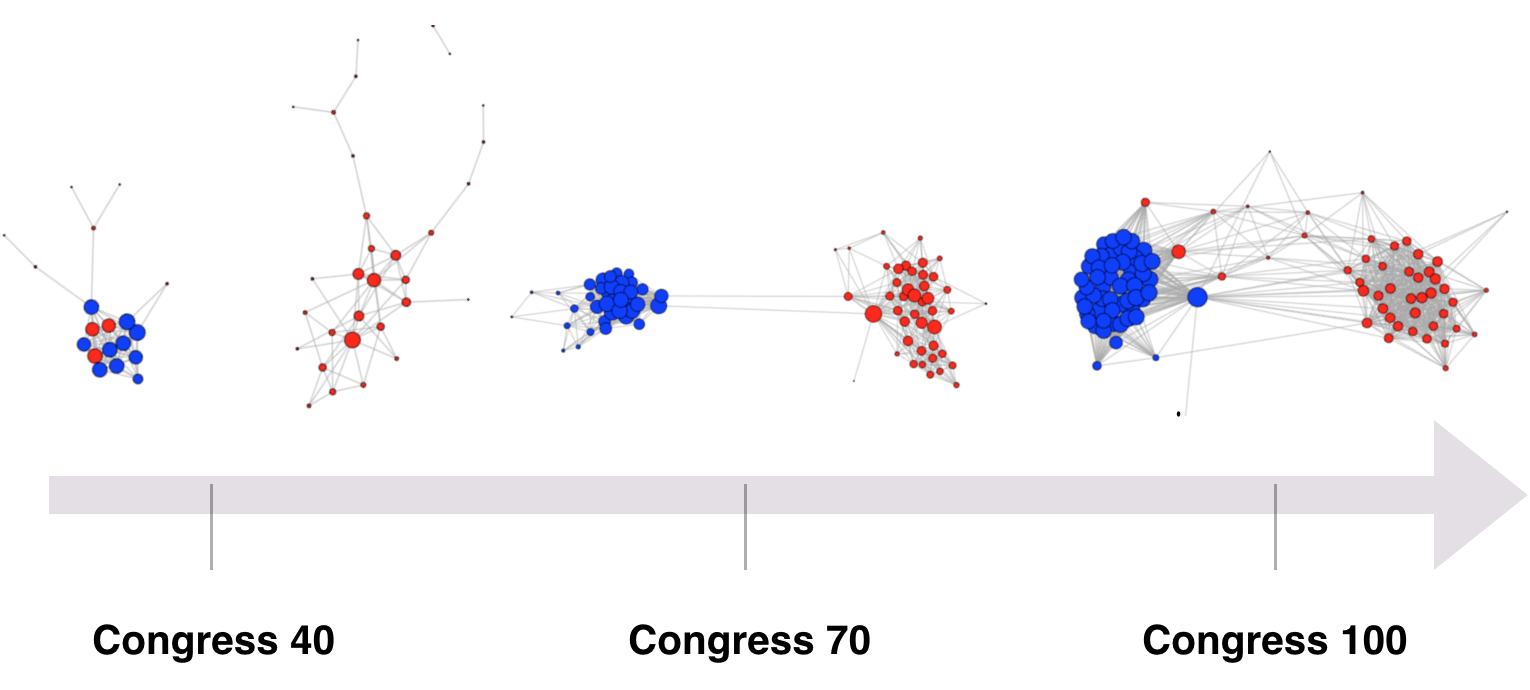}
	\caption{An illustration of the 40th, 70th, and 100th Senate networks in the U.S. Senate voting network. Each network was drawn using the Fruchterman Reingold layout. Nodes are colored according to political affiliation, where red represents Republican and blue represents Democratic affiliation. \label{fig:senate_pic}}
\end{figure}

\begin{figure}[ht]
	\centering
	\includegraphics[width = \textwidth, trim = 0cm .15cm 0cm 0cm, clip = TRUE]{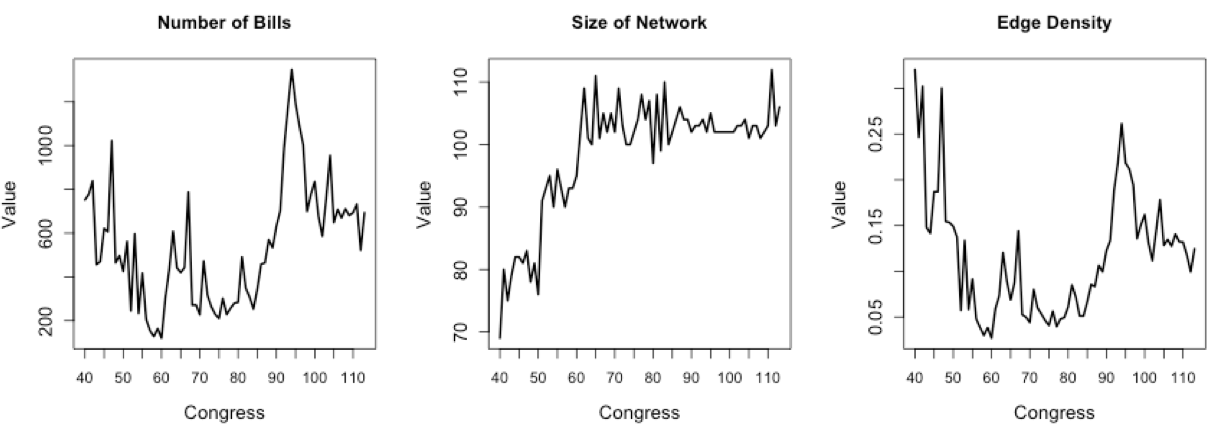}
	\caption{Features of the dynamic U.S. Senate voting network. \label{fig:descriptive}}
\end{figure}

We generated a dynamic network to model the co-voting patterns among U.S. Senators in the following manner. We first collected the roll call voting data for each Congress from \url{http://voteview.com}. This data set contains the voting decision (either yay, nay, or abstain) of each Senator for every bill submitted to the Senate. For each Congress, we model the Senators in that Congress as the collection of nodes. Binary edges are placed between two Senators if they vote concurrently (either both yay or both nay) for at least $75\%$ of the total number of bills on which either of them voted. Three of the networks that we analyze are shown in Figure \ref{fig:senate_pic}. This figure illustrates the tendency of the Senators to vote according to his or her own party affiliation. We summarize the number of Senators, number of bills, and the total number of edges in each Congress in Figure \ref{fig:descriptive}.

\subsection{Results}

\begin{figure}[H]
	\centering
	{\bf Shewhart Control Charts}
		\includegraphics[width = \textwidth, trim = 0cm 0cm 0cm 0cm, clip = TRUE]{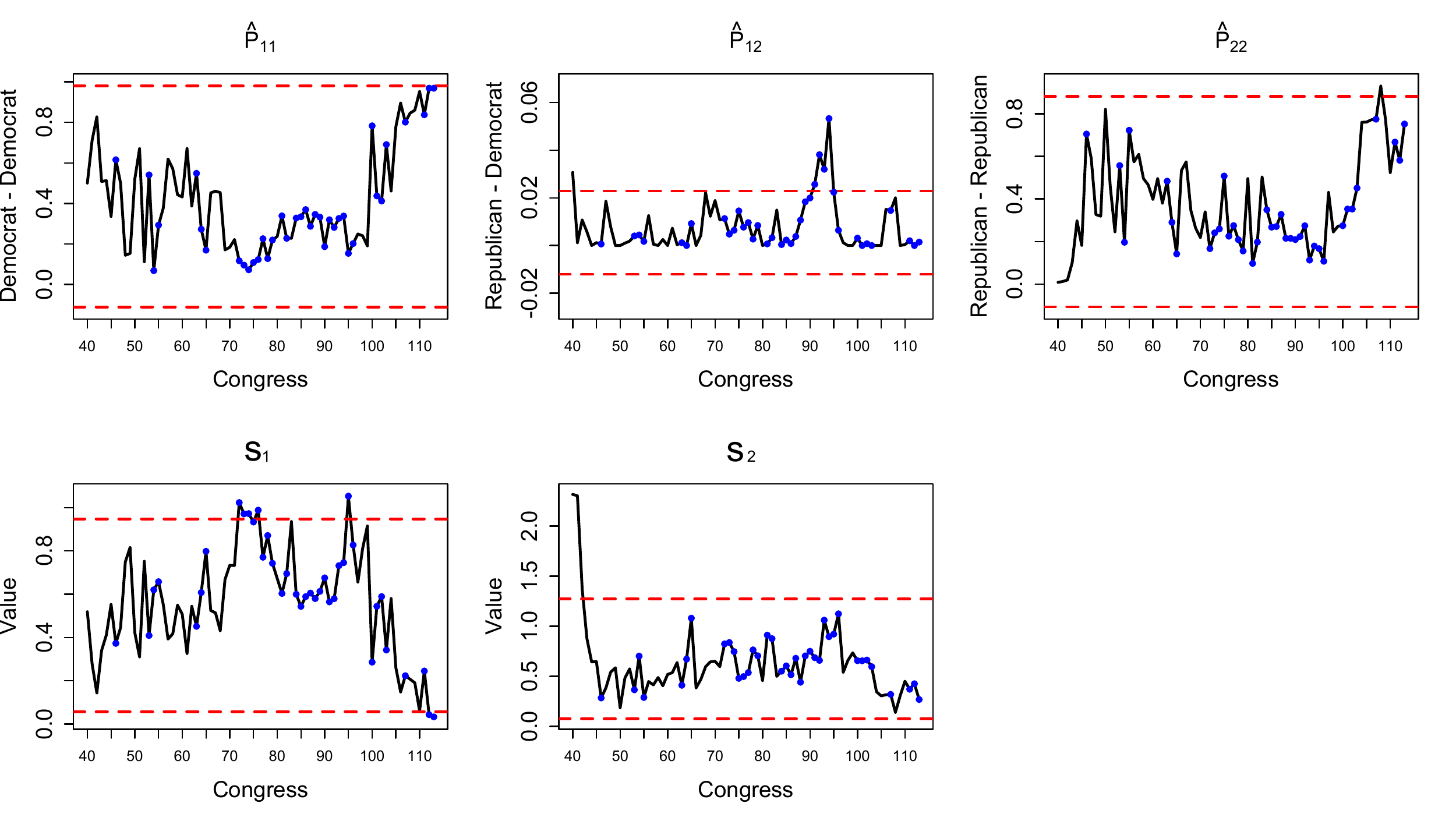} 
	{\bf EWMA Control Charts}
	\includegraphics[width = \textwidth, trim = 0cm 0cm 0cm 0cm, clip = TRUE]{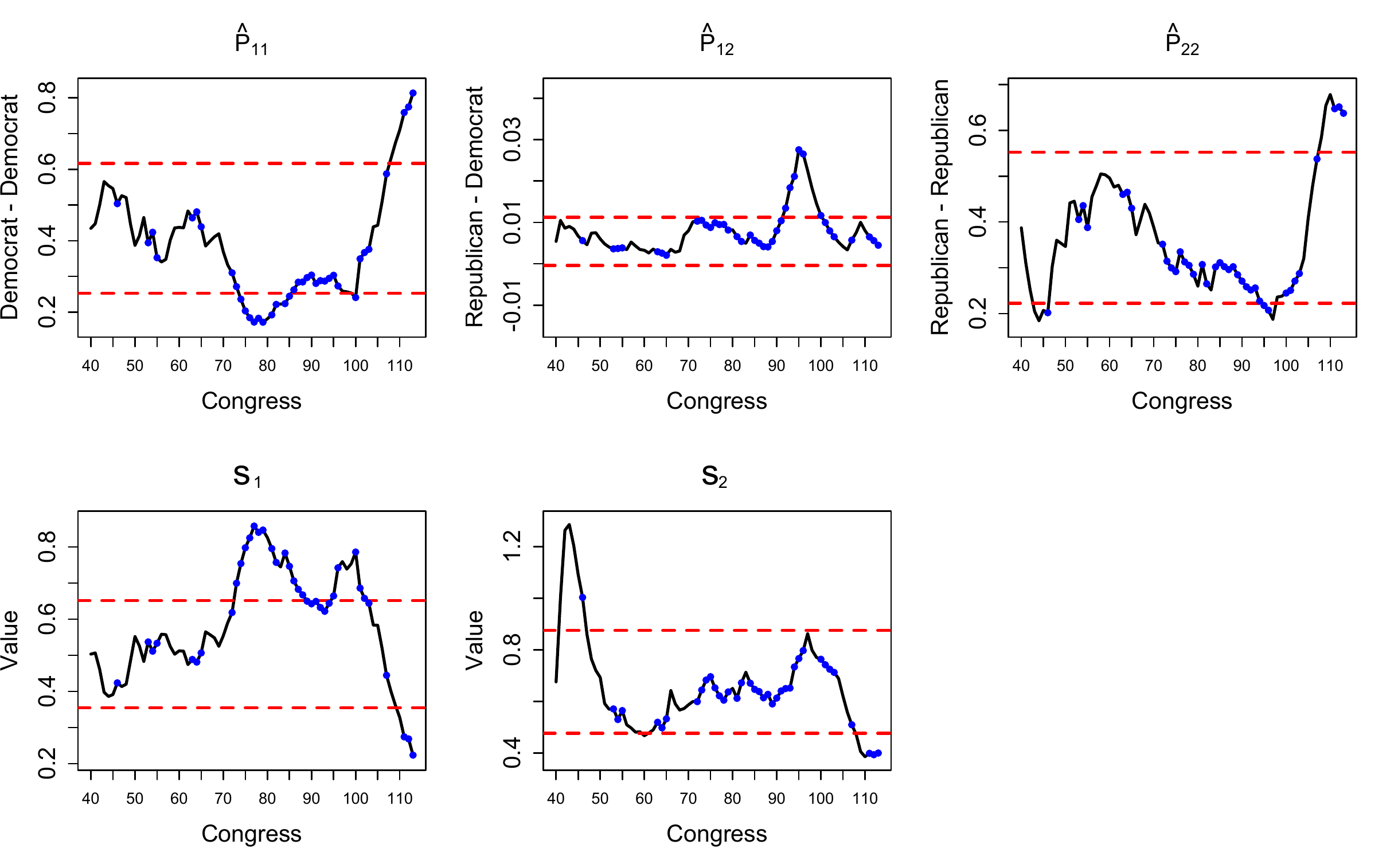} 
	\caption{Shewhart (top) and EWMA (bottom) control charts for each of the DCSBM statistics for the dynamic voting network of the U.S. Senate, when using a Phase I size of 25. The red dashed lines represent the upper and lower control limits for the Shewhart chart. Blue dots represent Congresses for which Democrats held a majority in Senate. These control charts illustrate a recent schism among Republican and Democratic voting patterns in the Senate as well as an era of political cohesion during the ''Rockefeller Republican'' era.  \label{fig:political_control1}}
\end{figure}

To analyze political polarization, we applied the DCSBM surveillance strategy with Shewhart and EWMA control charts to this dynamic network. Since node labels across graphs are not registered, i.e., nodes do not represent the same Senators across time, estimating the community labels using the spectral clustering strategy mentioned in Section \ref{sec:community_est} is not appropriate. As we are interested in understanding political polarization, we instead set the community labels at time $t$ according to the political affiliation of each Senator (1 for Democrat and 2 for Republican). We set the Phase I size to be $m = 25$ and compute the Shewhart and EWMA control charts for the estimators $\{\widehat{P}_{q,r}, s_q: q, r = 1, 2\}$. For the EWMA chart, we calculated the control limits in (\ref{eq:EWMA}) and set $\lambda = 0.2$. Estimation of the DCSBM and surveillance took approximately two minutes to run on this data set using R software on a laptop with a 2.6 GHz Intel Core i5 processor. The Shewhart and EWMA control charts are shown in Figure \ref{fig:political_control1}. 

The control charts in Figure \ref{fig:political_control1} reveal three interesting and relevant features about the U.S. Senate voting patterns. First, both the Shewhart and EWMA control charts signal \emph{large} values of $\widehat{P}_{1,2}$ from Congress 91 (1969 - 1971) to Congress 94 (1975 - 1977). This finding suggests that Republicans and Democrats tended to vote concurrently more often than expected during this period of time. Furthermore, the EWMA control chart signals large values of $s_1$ during this time period. This suggests that the voting propensity of the Democratic party during this time is significantly more variable than expected. Interestingly, this time frame lies at the second half of the so-called ``Rockefeller Republican'' era, which lasted from 1960 to 1980. During this era, many Republican Senators had moderate views that reflected the ideals of the governor of New York, Nelson Rockefeller \citep{rae1989decline, smith2014his}. The Rockefeller Republicans were strong supporters of the civil rights movement, including the Civil Rights Act of 1968, and held especially moderate fiscal views under the Presidency of Richard Nixon (93rd Congress). Notably, this general cohesion among parties - marked by large values of $\widehat{P}_{1,2}$ in the control charts - ended in Congress 94. This Congress coincides with the end of Nelson Rockefeller's role as Vice President of the United States in 1977. To the best of our knowledge, this is the first work to identify this political era using Senatorial co-voting data. 

Next, the EWMA control charts for $\widehat{P}_{1,1}$ and $\widehat{P}_{2,2}$ signal \emph{large} values at Congress 104. This suggests that the intra-party co-voting propensities for both the Democratic and Republican parties became exceedingly large at that time. This finding supports the theory of recent polarization of the parties at the beginning of Bill Clinton's first term as President (Congress 103). According to \citet{moody2013portrait}, this time period marked an important transition at which conservative Democrats and liberal Republicans joined majority-party coalitions in both Congress 103 (Democratic majority) and Congress 104 (Republican majority). This transition left the middle ground between parties empty, which may have lead to an enduring polarization. These results also coincide with the findings of \citet{roy2014change}. The Shewhart control chart did not as clearly signal this change; however, in each of the charts there is an increasing trend beginning in Congress 100. We note that the Shewhart lower control limit for $\widehat{P}_{1,1}$, $\widehat{P}_{1,2}$, and $\widehat{P}_{2,2}$ is less than zero. This indicates that the variability of these values in Phase I was too large to construct tight control limits. As Shewhart charts are better suited for large sudden changes, it is expected that these charts will identify this polarization change later than the EWMA chart, as shown. 

Finally, we see from the EWMA control charts for $s_1$ and $s_2$ signals a significantly small value of these statistics at Congress 105. This suggests that the variability of total interaction of the Senators steadily and significantly reduced during this period. This finding complements the polarization theory described above, and suggests that since Congress 105, each U.S. Senator tends to vote according to his or her party, regardless of the bill.

\section{Simulation Study}\label{sec:numerical_study}
In this section, we investigate the detection of structural changes in a network $\widehat{\bm{G}}(n,T) = \{\widehat{G}_1, \ldots, \widehat{G}_T\}$ generated under a dynamic DCSBM. We consider local and global changes in the network as parameterized by changes in $P \rightarrow P^*$, $\bm{\delta} \rightarrow \bm{\delta}^*$, and $\bm{c} \rightarrow \bm{c}^*$ at time $t^*$. Note that we assume the community labels $\bm{c}$ are known and so we do not surveil the maximum likelihood estimates $\widehat{\bm{\pi}}$. Because these local and global changes are large and introduced suddenly, we use Shewhart control charts as the monitoring strategy.

In Section \ref{sec:illustrative} we evaluate this monitoring strategy on a collection of illustrative examples to gain an intuition of the DCSBM and the performance of the proposed methodology. In Section \ref{sec:runlength} we quantify the strengths and weaknesses of our method using an analysis of \emph{average run lengths} under a variety of simulated conditions. To evaluate the performance of our detection strategy, we altered the network size and the magnitude of the change being introduced. This simulation strategy can be readily used to assess the performance of any network surveillance method. 

\begin{table}
	\caption{A description of the changes introduced to the dynamic DCSBMs in our simulation study. \label{table:simulation_example}}
	\centering
	\begin{tabular}{c | c | l}
		{\bf Simulation} & {\bf Change} & {\bf Description} \\ \hline
		1 & $P^*_{1,1} = P_{1,1} + {\epsilon}$ & local outbreak in community 1\\
		2 & $P^*_{i,j} = P_{i,j} + {\epsilon}$ & global outbreak $(i=1,2$, $j = 1,2)$\\
		3 & $\delta^*_1 = \delta_1 + {\tau}$ & local variability increase in community 1\\
		4 & $\delta^*_i = \delta_i + {\tau}$ & global variability increase $(i=1,2)$\\
		5 & $\bm{c} \rightarrow \bm{c}^*$ & merge communities \\
		6 & $\bm{c} \rightarrow \bm{c}^*$ & split community 1 into 2 communities\\ \hline
	\end{tabular}
	
\end{table}

\subsection{Illustrative Examples}\label{sec:illustrative}

We begin our simulation study by demonstrating the Shewhart control charts on a collection of six dynamic networks, each of which reflects a different structural change at time $t^*$. We investigate changes in the mean and variance of interaction rate, both locally and globally, as well as changes in community structure. For each simulation, we generated a dynamic network according to (\ref{eq:sims}) with $n = 50$ nodes, $k = 2$ equally sized communities, $T = 50$ time points, and a change implemented at time $t^* = 30$. We use the first $m = 25$ simulated networks for Phase I, and implemented the {Shewhart} control chart for the statistics $\{\widehat{P}_{q,r}, s_{q}: q,r = 1, 2\}$ using the surveillance strategy described in Section \ref{sec:strategy}. In all six simulations, we set 

$$P = \begin{pmatrix} 0.2 & 0.1 \\ 0.1 & 0.2 \end{pmatrix}, \hskip 1pc \delta_r \equiv 0.5 \hskip .5pc \text{for $r = 1, 2$}.$$ 

\begin{figure}
	\centering
	\begin{tabular}{c}
		{\bf Simulation 1}\\
		\includegraphics[width = \textwidth]{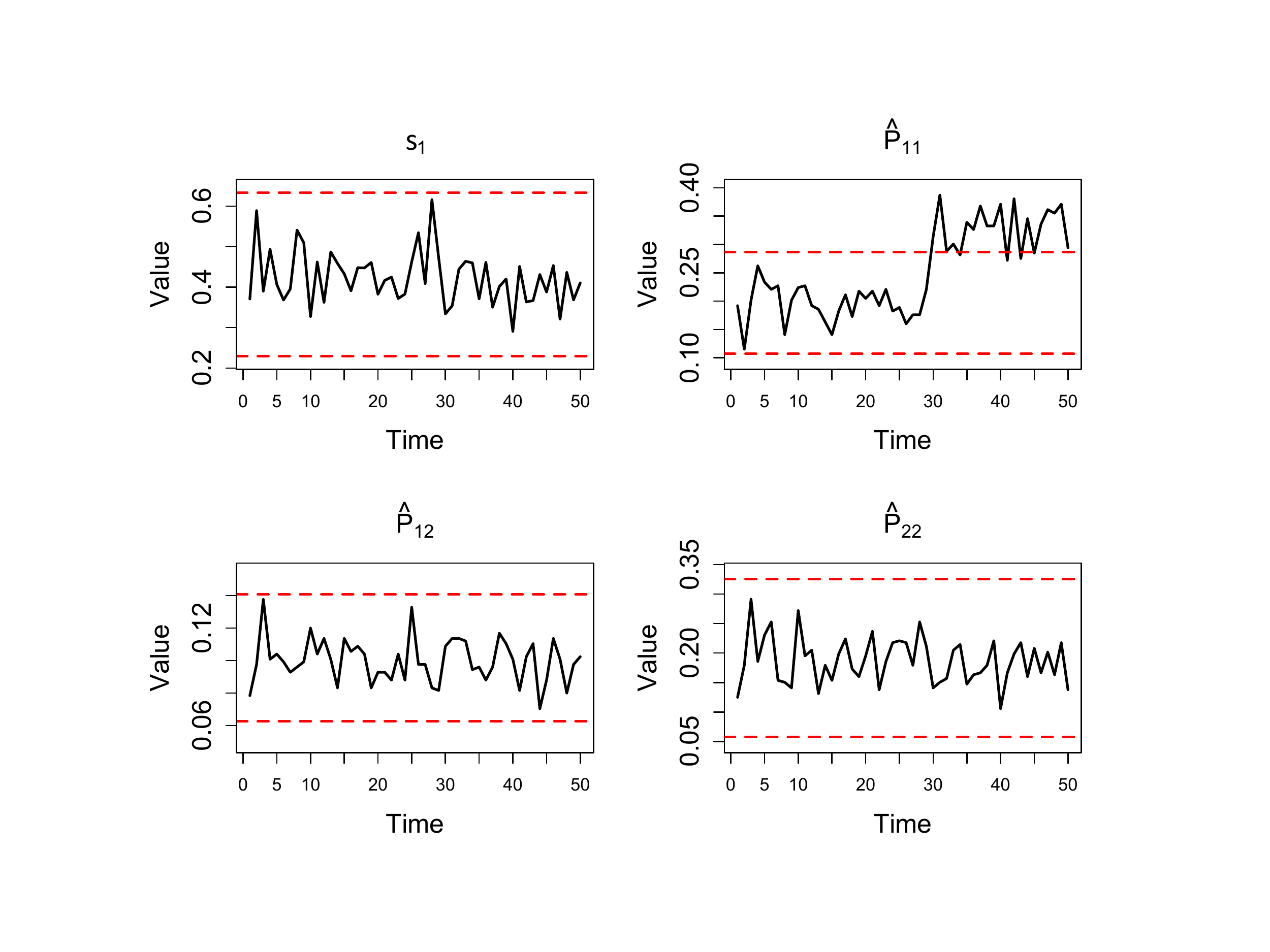} \\
		{\bf Simulation 2}\\
		\includegraphics[width = .75\textwidth, trim = 0.025cm 0cm 0cm 0cm, clip = TRUE]{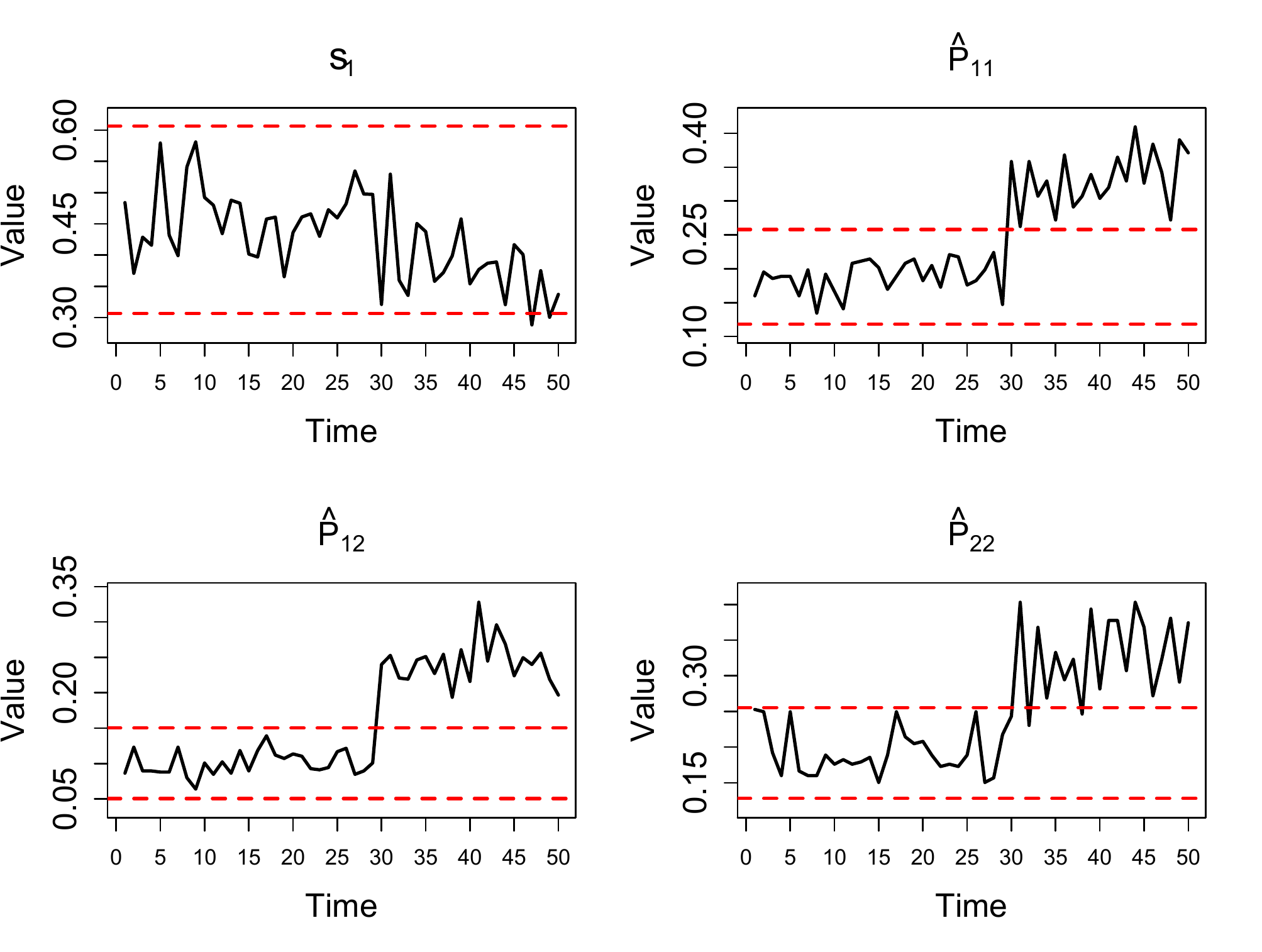}
	\end{tabular}
	\caption{Shewhart control charts for the dynamic networks generated for simulations 1 and 2 estimated using the first 25 networks.  \label{fig:sim12}}
\end{figure}

\begin{figure}
	\centering
	\begin{tabular}{c}
		{\bf Simulation 3}\\
		\includegraphics[width = .72\textwidth]{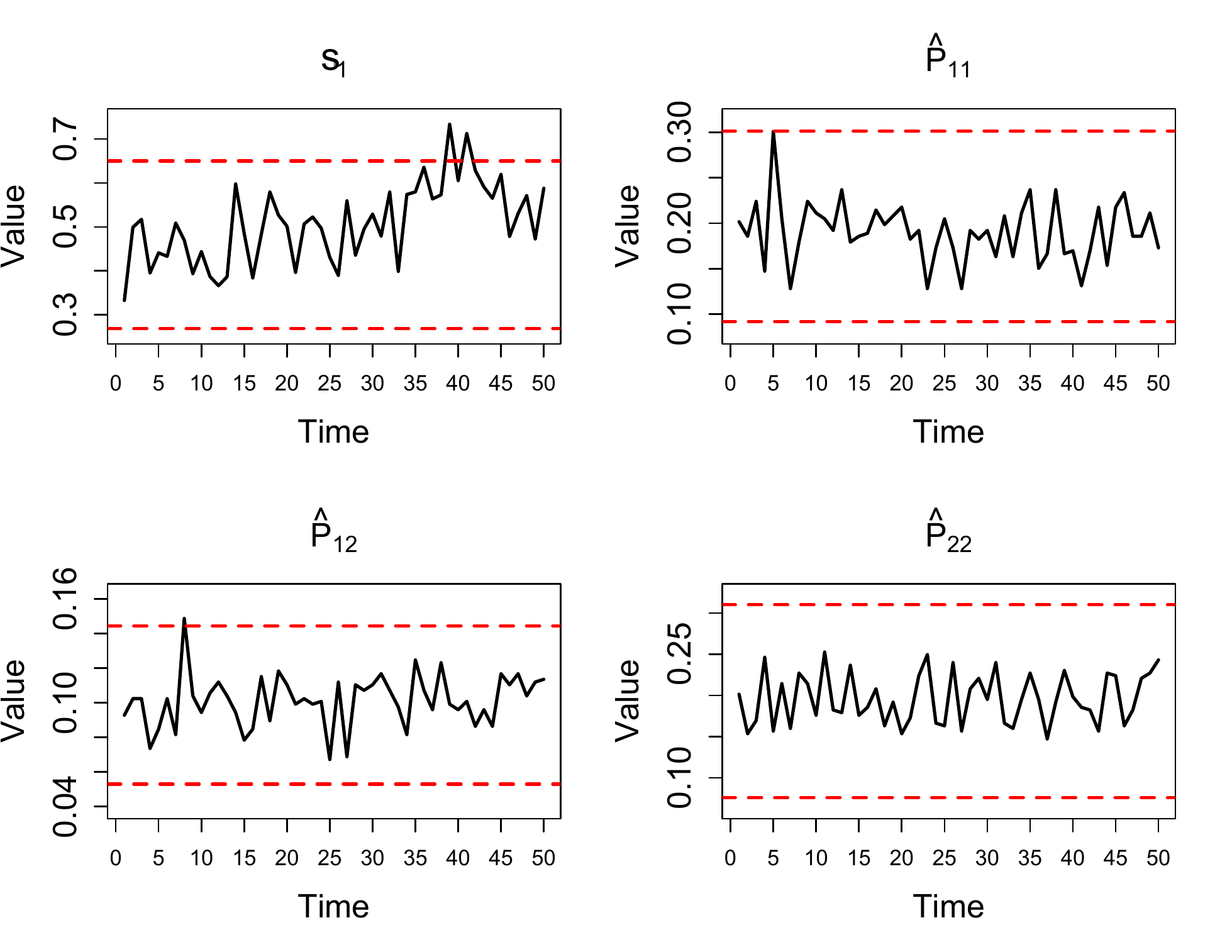} \\
		{\bf Simulation 4}\\
		\includegraphics[width = .72\textwidth, trim = 0cm 0cm 0cm 0cm, clip = TRUE]{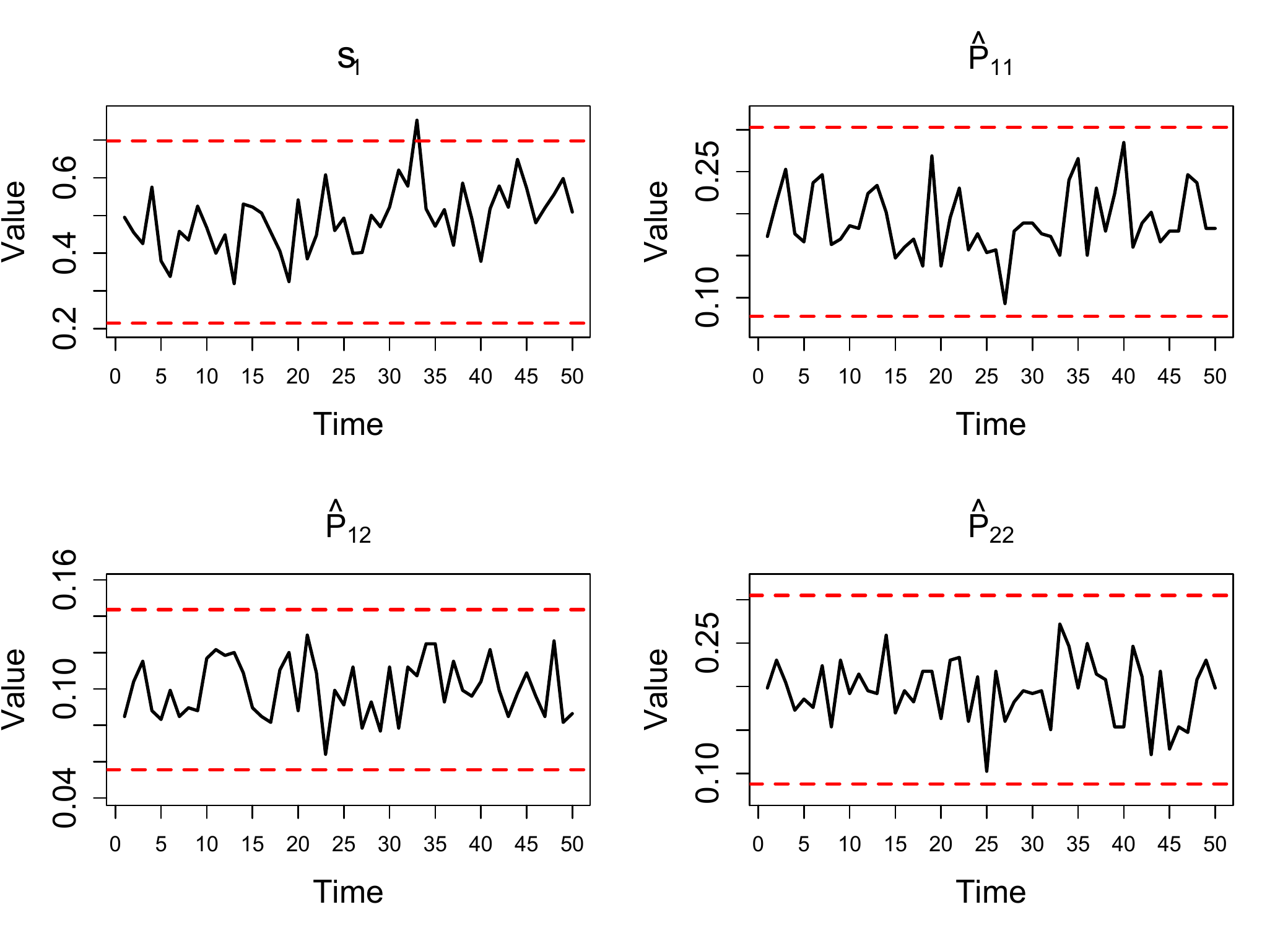}
	\end{tabular}
	\caption{Shewhart control charts for the dynamic networks generated for simulations 3 and 4 estimated using the first 25 networks.\label{fig:sim34}}
\end{figure}

\begin{figure}
	\centering
	\begin{tabular}{c}
		{\bf Simulation 5}\\
		\includegraphics[width = .72\textwidth]{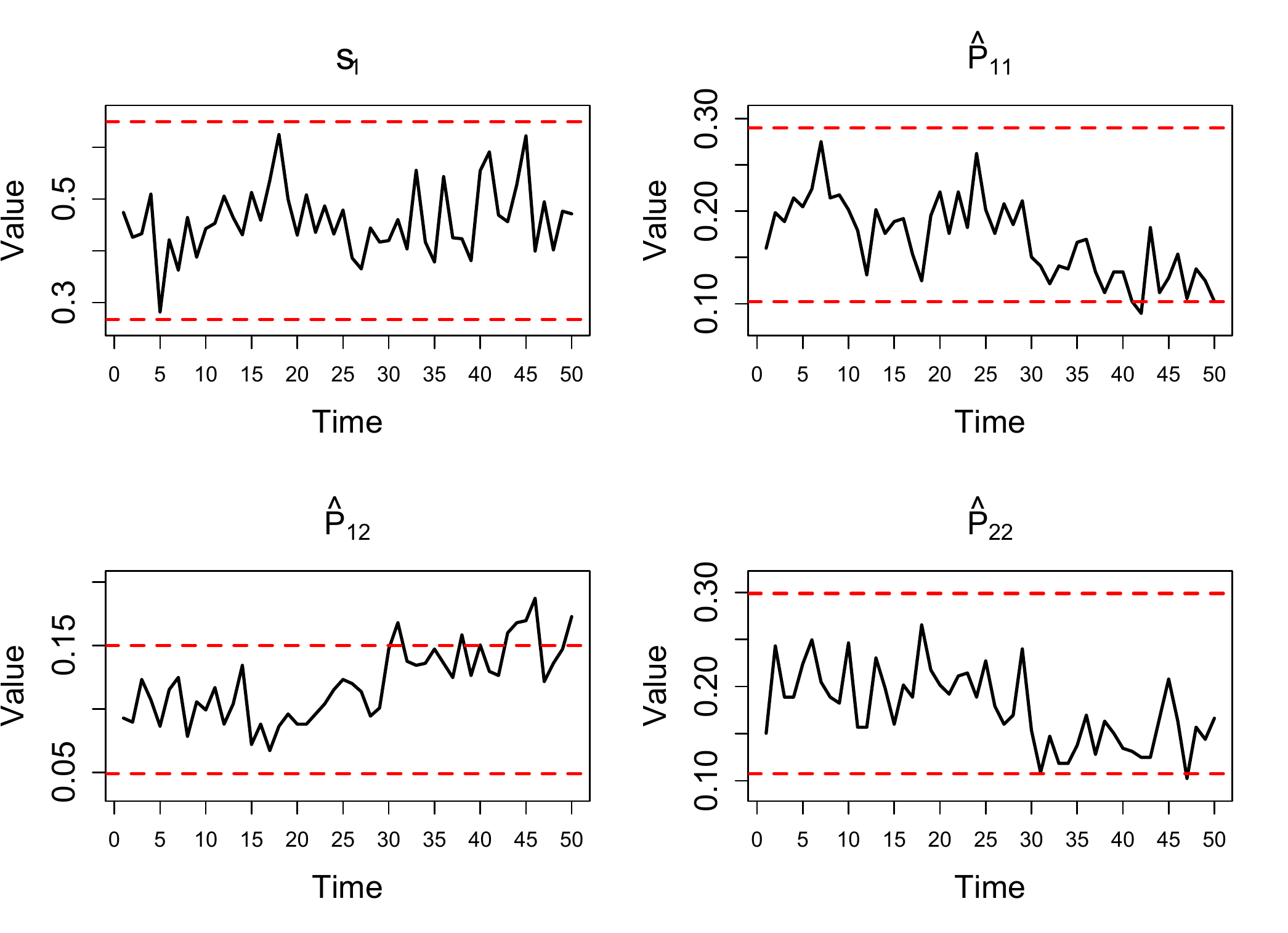} \\
		{\bf Simulation 6}\\
		\includegraphics[width = .72\textwidth]{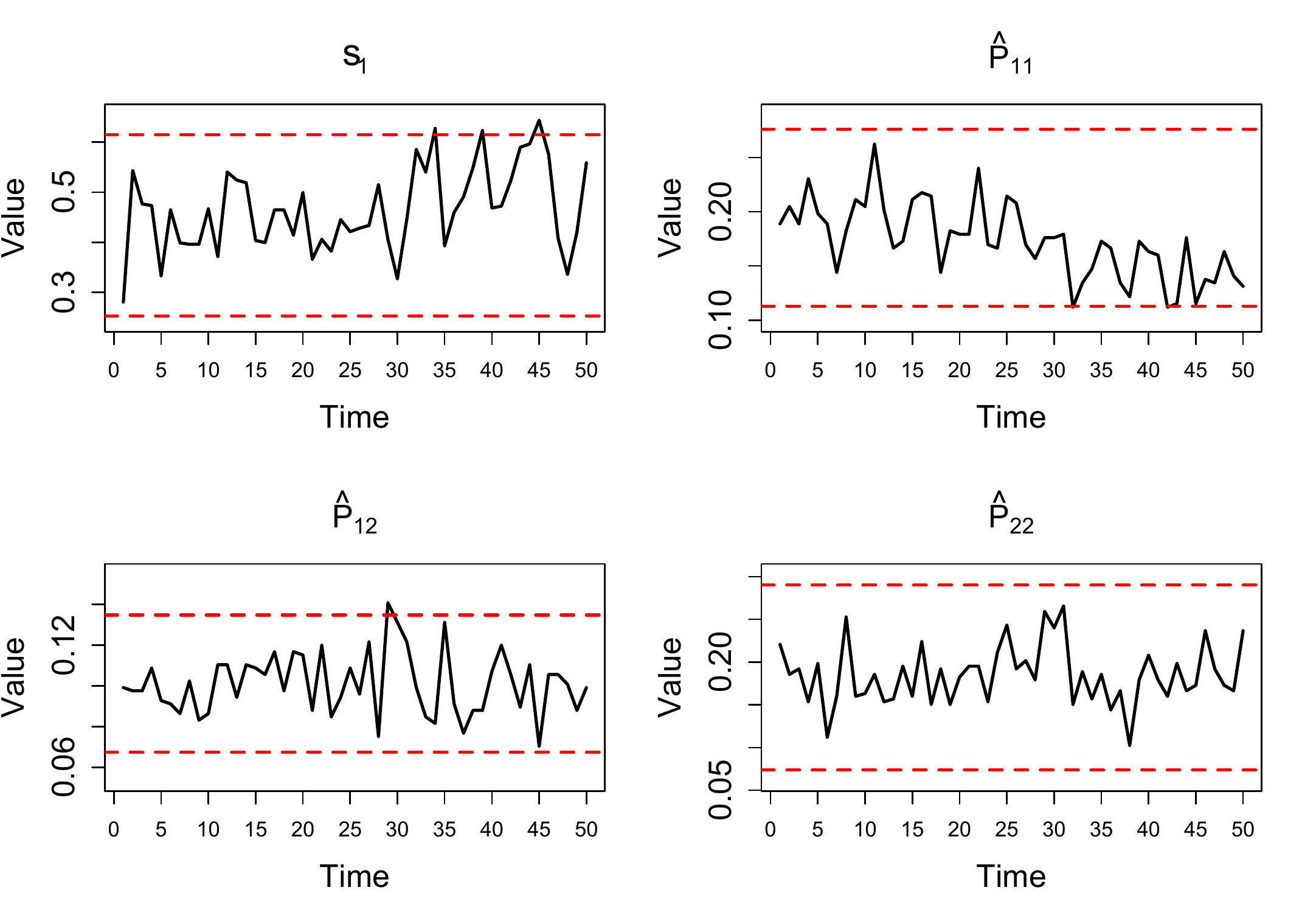}
	\end{tabular}
	\caption{Shewhart control charts for the dynamic networks generated for simulations 5 and 6 estimated using the first 25 networks. \label{fig:sim56}}
\end{figure}

Control charts are shown for each simulation in Figures \ref{fig:sim12}, \ref{fig:sim34}, and \ref{fig:sim56}. Below, we describe the six simulated networks and the results of our monitoring plan. To conserve space we do not present charts for $s_{2}$, and instead describe them qualitatively where appropriate. The implemented changes for each simulation are described in Table 1.

\noindent \underline{Simulations 1 - 2: Mean Interaction Rate Changes}
\vskip .5pc


In the first two simulations, we monitor changes in the mean interaction rates in the network. In simulation 1, we introduce a local mean interaction outbreak in community 1 by setting $P^*_{1,1} = P_{1,1} + \epsilon$ {with $\epsilon=0.10$}. The top of Figure \ref{fig:sim12} reveals that the control chart for $\widehat{P}_{1,1}$ efficiently signals a change at time 30; whereas, all other statistics remain in control over the entire time interval. In simulation 2, we introduce a global mean interaction outbreak by increasing all entries of $P$ by $\epsilon=0.10$. In this case, the probability estimates $\widehat{P}_{1,1}$, $\widehat{P}_{1,2}$ and $\widehat{P}_{2,2}$ all lead to a signal for a change at time 30, and $s_{1}$ and $s_{2}$ remain in control, though the chart for $s_{2}$ is not shown here. We note that $\widehat{P}_{1,2}$ appears to signal the most dramatic change. This is due to the fact that the signal to noise ratio introduced by increasing the overall interaction rate in the network is highest for the inter community interactions. 

\newpage
\noindent \underline{Simulations 3 - 4: Variance of Interaction Rate Changes}
\vskip .5pc

Next we monitor changes in the variation of the interaction rate in the simulated network. In simulation 3 we increase $\delta_1$ {by $\tau=0.25$, which} results in a change in the variability of interaction in community 1. The top of Figure \ref{fig:sim34} reveals that this change is indeed signalled by the $s_{1}$ chart near $t = 30$. We expect the reaction of the chart, and hence the signal delay, to depend on the magnitude of change. We investigate this further in the next section. In simulation 4 we simulated a global change in $\bm{\delta} = (\delta_1, \delta_2)$, which increases the variability of interactions among all nodes. {In this case $\delta_1$ and $\delta_2$ are both increased by $\tau=0.25$.} The bottom of Figure \ref{fig:sim34} reveals that $s_1$ signals the change almost immediately. Although not shown here, the control chart for $s_2$ behaves similarly. Importantly, the connection probability estimates remain in control in these simulations suggesting, as desired, that the mean interaction rate in the network does not change. 

\vskip .5pc
\noindent \underline{Simulations 5 - 6: Change in Community Structure}
\vskip .5pc
In simulations 5 and 6, we consider two common changes in community structure: merging and splitting of communities. In simulation 5, we simulate networks with two equally sized communities up to time $t^* = 30$. At time $t^*$, we then merge the two communities into one and set the connection value to the average of the former connection probabilities, that is $P^* = 0.15$. Structurally, this change results in an increase of $\widehat{P}_{1,2}$ by 0.05 and a decrease in $\widehat{P}_{1,1}$ and $\widehat{P}_{2,2}$ by 0.05. Our control charts from Figure \ref{fig:sim56} detect this trend, and we see that the change is appropriately detected using $\widehat{P}_{1,2}$. Although we witness a decrease in $\widehat{P}_{1,1}$ and $\widehat{P}_{2,2}$, the control chart does not {signal} a change immediately. Because this change is relatively small, it would be better detected by EWMA control charts for $\widehat{P}_{1,1}$ and $\widehat{P}_{2,2}$.

In simulation 6, we once again begin with two equally sized communities. At time $t^* = 30$, we split community 1 into two communities of size 12 and 13, respectively. For the three communities after time $t^*$, we fix $P_{i,i} = 0.20$ and $P_{i,j} = 0.10$ as before. Structurally such a change will be reflected by an overall decrease in $\widehat{P}_{1,1}$. We see this trend in the chart in the bottom of Figure \ref{fig:sim56}; however, the change was not identified until time $t=42$, where $\widehat{P}_{1,1}$ went below the control limits. We expect that this type of change will be more readily detected in larger networks and in networks where the split community is large. We investigate this further in the next section.

{
\subsection{Average Run Length Analysis}\label{sec:runlength}

For each scenario described in Table 1, we evaluated our monitoring methodology by simulating the situation 1000 times. On each of these 1000 simulated runs, we calculated the number of networks until the control chart detects a change, i.e., the run length, and we then estimate the average run length (ARL) from these 1000 simulations. Because $\mu$ and $\sigma$ are estimated from Phase I, there will be practitioner-to-practioner sampling differences in observed ARL values, which is the basis for an ARL distribution. Thus the average run lengths we report are estimates of the mean of this distribution, which we refer to as the average ARL (AARL) as in \citet{saleh2015arl}. This AARL is the basis upon which different surveillance methods can be compared. In what follows, we describe the performance of the surveillance technique discussed in the previous two sections.

In each of the scenerios discussed below we assume the same initial form of $P$ and $\bm{\delta}$ as discussed in the previous section, with $n=100$ nodes in each network. We investigated the performance of the method with $m = 25$, $m = 50$ and $m = 1000$ Phase I samples. In all cases we implemented the appropriate change at time $t^*=25$ in Phase II and thereafter generated as many networks required to observe the first signal on each control chart. Here we investigate the performance of control charts for $\widehat{P}_{1,1}$, $\widehat{P}_{1,2}$, $\widehat{P}_{2,2}$, and $s$ which is a pooled estimate of the standard deviation of $\widehat{\bm{\theta}}$ based on $s_1$ and $s_2$ since we assume $\delta_{1} = \delta_{2}$ in Phase I.

We found comparable performance of our surveillance technique under Phase I sizes of $m = 25, m = 50$ and $m=1000$.  However, as \citet{saleh2015arl} indicate, it is unwise to guarantee specific ARL values when the control chart parameters are estimated from small sample sizes. As such, we present the results of the $m = 1000$ case here, and provide the results for the $m = 25$ and $m=50$ cases in the Appendix. Note that when $m=1000$, we gain insight into the performance of the methodology under favorable conditions (i.e., when information about each statistic's distribution is ample). 

\vskip .5pc
\noindent \underline{Simulation 0: No Change}
\vskip .5pc

We begin by considering the performance of the methodology when no structural change has occurred. Doing so allows us to quanitify the prevalence of \emph{false alarms}, i.e., when the control chart incorrectly indicates a change has occurred. The AARLs associated with the control charts for $s$, $\widehat{P}_{1,1}$, $\widehat{P}_{1,2}$, and $\widehat{P}_{2,2}$ are shown in Table \ref{table:ARLm1000}. Although there will be variation in in-control ARLs, the large AARL values shown in the Simulation 0 row are reassuring; they indicate that false alarms are not expected to occur until hundreds of ``in-control'' networks have been observed. When structural changes have occurred, we expect much smaller AARLs to be associated with at least one of the four control charts. We discuss these scenarios below.


\vskip .5pc
\noindent \underline{Simulations 1 - 2: Mean Interaction Rate Changes}
\vskip .5pc
We quantify the method's ability to detect \emph{local} changes in $P$, specifically in community 1, by adding $\epsilon=0.01, 0.05, 0.10$ to $P_{1,1}$. As mentioned previously, such a change is expected to be detected on the $\widehat{P}_{1,1}$ control chart. The Simulation 1 AARLs in the $\widehat{P}_{1,1}$ column of Table \ref{table:ARLm1000} indicate that this is indeed the case; on average we expect the $\widehat{P}_{1,1}$ control chart to detect such a change in roughly ten networks for moderate sized changes in $P_{1,1}$, and roughly two networks for large changes. On the other hand, the large AARL values for the other three statistics indicate that none of them is likely to detect this change, as desired.

We similarly quantify the method's ability to detect \emph{global} changes in $P$ by adding $\epsilon=0.01, 0.05, 0.10$ to each $P_{i,j}$. In this situation, we expect all entries of $\widehat{P}$ to signal a change. The Simulation 2 AARLs in the $\widehat{P}_{1,1}$, $\widehat{P}_{1,2}$, and $\widehat{P}_{2,2}$ columns of Table \ref{table:ARLm1000} support this hypothesis. As expected, we see that the $\widehat{P}_{1,2}$ control chart signals this change fastest since $\epsilon$ is much larger relative to $P_{1,2}$ than it is to $P_{1,1}$ and $P_{2,2}$. 


\vskip .5pc
\noindent \underline{Simulations 3 - 4: Variance of Interaction Rate Changes}
\vskip .5pc

We introduced \emph{local} changes in interaction variability among the nodes in community 1 by adding $\tau=0.05, 0.10, 0.25$ to $\delta_1$, and we introduce \emph{global} changes in interaction variability among all nodes in the network by adding $\tau=0.05, 0.10, 0.25$ to each $\delta_j$, $j=1, 2$. In both cases, we expect the $s$ control chart to signal this change. The AARLs in the Simulation 3 and Simulation 4 rows of Table \ref{table:ARLm1000} support this claim. In particular, we can expect this control chart to detect global changes more quickly than local changes, and in both cases large changes will be detected more quickly than small changes. 


\vskip .5pc
\noindent \underline{Simulations 5 - 6: Change in Community Structure}
\vskip .5pc

As discussed in the previous section, Simulation 5 corresponds to the merging of communities. Since $P_{1,2}$ is most affected by this change, we expect the $\widehat{P}_{1,2}$ control chart to signal quickest. The AARLs in the ``Simulation 5'' row of Table \ref{table:ARLm1000} agree with this intuition; while, $\widehat{P}_{1,1}$ and $\widehat{P}_{2,2}$ tend to detect this change more quickly than $s$, the $\widehat{P}_{1,2}$ chart detects the change almost immediately. Interestingly, this result does not appear to depend on the size of the network. 


When community $j$ is split into two (equally sized) communities, the illustrative example in Section \ref{sec:illustrative} suggests that a control chart for $\widehat{P}_{j,j}$ should signal most quickly. The results in the Simulation 6 row of Table \ref{table:ARLm1000} substantiate this; when community 1 is split into two communities, the control for $\widehat{P}_{1,1}$ detects this more quickly than the other control charts, but perhaps not as quickly as a practitioner would like. This suggests that the proposed surveillance methodology may not be ideal for detecting community splitting, even though it is highy effective at detecting each of the other types of structural change considered. 


\begin{table}
	\caption{Average ARLs for Simulations in Section 7.2 when $m=1000$. \label{table:ARLm1000} }
	\centering
	\begin{tabular}{cclllll}
		{\bf Simulation} & \multicolumn{2}{c}{\bf Change} & {$s$} & {${\widehat{P}_{1,1}}$} & {${\widehat{P}_{1,2}}$} & {${\widehat{P}_{2,2}}$} \\ \hline
		0 & & none & 317.18 & 439.25 & 446.50 & 338.25 \\ \hline
		  & & $\epsilon = 0.01$ & 294.80 & 134.00 & 413.70 & 332.4\\
		1 & $P^*_{1,1} = P_{1,1} + \epsilon$ & $\epsilon = 0.05$ & 284.90 & 9.87   & 257.27 & 207.70\\
		  & & $\epsilon = 0.10$ & 524.40 & 2.23   & 289.90 & 325.90\\ \hline
		  & & $\epsilon = 0.01$ & 498.80 & 140.90 & 64.65  & 142.30\\
		2 & $P^*_{i,j} = P_{i,j} + \epsilon$ & $\epsilon = 0.05$ & 211.10 & 9.48   & 1.71   & 12.17\\
		  & & $\epsilon = 0.10$ & 93.30  & 2.01   & 1.01   & 2.28\\ \hline		  
	      & &$\tau = 0.05$  & 106.51 & 221.40 & 260.10 & 202.70\\
	    3 & $\delta^*_1 = \delta_1 + \tau$ & $\tau = 0.10$  & 115.70 & 152.33 & 305.29 & 544.60\\
	      & & $\tau = 0.25$  & 18.81  & 63.35  & 107.20 & 431.00\\ \hline
  	      & & $\tau = 0.05$  & 93.58  & 232.30 & 246.10 & 216.10\\
  	    4 & $\delta^*_i = \delta_i + \tau$ & $\tau = 0.10$  & 36.33  & 142.00 & 185.94 & 218.50\\
  	      & &$\tau = 0.25$  & 4.94   & 52.88  & 92.23  & 53.87\\ \hline
    	  & &$n = 50$          & 327.60 & 74.97  & 1.64   & 40.81\\
    	5 & Merge comm. &$n = 100$         & 247.00 & 39.79  & 1.66   & 27.61\\
    	  & & $n = 500$         & 72.70  & 37.56  & 1.61   & 37.32\\ \hline
      	  & & $n = 50$          & 152.10 & 32.88  & 168.30 & 427.80\\
      	6 & Split comm. &$n = 100$         & 127.50 & 33.90  & 313.39 & 426.20\\
      	  & & $n = 500$         & 72.70  & 33.37  & 315.50 & 446.50\\ \hline
	\end{tabular}
\end{table}


\section{Discussion}\label{sec:discussion}
In this paper we have illustrated the utility of the dynamic degree corrected stochastic block model (DCSBM) in modeling and simulating realistic dynamic networks with local and global structural changes. Our proposed model is flexible, and can capture both degree heterogeneity and community structure in networks, two important features that are common in social and biological networks. We proposed a fast and effective monitoring methodology based on the surveillance of maximum likelihood estimates from the DCSBM using Shewhart and EWMA control charts for individuals. When applying our method to the U.S. Senate co-voting network, we were able to identify relevant and significant changes in the bipartisan nature of the U.S. Congress.
Our analysis reveals that the dynamic DCSBM can effectively model a variety of dynamic networks with structural changes, and that our proposed surveillance strategy can detect relevant changes in a real dynamic system. 

Our proposed monitoring strategy establishes one practically useful technique among a general family of methods for surveillance. Our framework relies on two components: a parametric dynamic random graph model for modeling the features of the graph, and a control chart from statistical process monitoring for the detection of changes in the parameters. We considered a dynamic DCSBM random graph model and the Shewhart and EWMA control charts for surveillance. This serves only as a first step in understanding the utility of our proposed surveillance strategy. In future work, it would be useful to explore the use of other parametric random graph models and control charts and to assess the advantages and disadvantages of each strategy. In particular, future work will explore the utility of dynamic latent space models like that discussed in \cite{sewell2015latent} as well as dynamic exponential random graph models like the TERGM family described in \cite{hanneke2010discrete}.

Our current surveillance framework requires the surveillance of on the order of ${k \choose 2}$ statistics, where $k$ is the number of communities in the network. If the number of communities is large, e.g., if $k = \mathcal{O}(n)$, our proposed surveillance strategy will become cumbersome and may suffer from multiple testing issues. For this reason, an important next step is to develop a surveillance methodology that is not limited by the number of nodes or communities in the network. For example, one could develop a formal likelihood ratio test for the DCSBM from one time point to the next. At every time point in Phase II, the likelihood ratio test statistic could be plotted on a control chart whose control limits are based on the exact or an approximate distribution of the statistic. The development of a likelihood ratio test for this network model is an important, but difficult, problem. \citet{yan2014model} provides some intuition for how to proceed here, but more work needs to be done.

Finally, the majority of contemporary surveillance methodologies are based on the assumption that the observed dynamic graph is unweighted. As a consequence, model-based approaches generally model the existence of an edge as a Bernoulli random variable and often rely on some thresholding technique to binarize count data. The DCSBM flexibly models the edge weight associated with each edge using a Poisson random variable. Thus, one can utilize the DCSBM to investigate and quantify the loss of information when count data is thresholded to binary outcomes.

\appendix
\section*{Appendix}\label{sec:appendix}

\subsection*{Appendix A: Relationship of DCSBM with other Random Graph Models}\label{sec:relatedmodels}
The DCSBM generalizes several families of well-studied and widely-applied random graph models. For the convenience of the reader, we describe three important families of random graph models and their relationship with the DCSBM below. The analysis of random graphs has a rich history, and a variety of models have been developed and used in a wide range of applications. \citet{goldenberg2010survey} and \citet{fienberg2012brief} provide two recent surveys of random graph models, and \citet{durrett2007random} provides a book level treatment of the topic. We refer the reader to these references and those mentioned below for more details about random graph theory and its application.

\begin{itemize}
	\item {\bf Stochastic block model}: When $\theta_u \equiv 1$ for all $u \in [n]$ and $P_{r,s} \in (0,1)$ for all $r,s \in [k]$, the degree corrected stochastic block model reduces to the (non-degree corrected) stochastic block model from \citet{holland1983stochastic, snijders1997estimation, nowicki2001estimation}. In this special case connection probabilities are fully described by the $k \times k$ probability matrix $P$. In this random graph, vertices in the same community are treated as {\it stochastically equivalent} in the sense that vertices of the same community have the same degree propensity. 
	\item {\bf \erdos\hskip -.05pc($p$)}: Suppose that $\theta_u \equiv 1$ and that $P_{r, s} \equiv p \in (0,1)$ for all $r, s \in [k]$. Then the DCSBM reduces to the \erdos random graph model with probability parameter $p$ \citep{erdos1960evolution}. The \erdos random graph model treats edges as independent and identically distributed random variables with connection probability $p$. As a result, the model does not distinguish vertices of different communities. The \erdos random graph is often used as a null model to which significant network features can be detected through comparison. For example, the \erdos random graph plays an important role in community detection both as a means to identify communities \citep{newman2006modularity}, and as a means to analyze the theoretical properties of community detection algorithms \citep{bickel2011method}. 
	\item {\bf Chung-Lu model}: An important family of random graph models is the family of fixed degree random graphs. These models are used to characterize the degree heterogeneity of an observed graph with degree sequence $\mathbf{d} = \{d(1), \ldots, d(n)\}$. A fixed degree random graph is a probability measure on the family of undirected graphs that have degree sequence $\mathbf{d}$. Important examples of fixed degree random graphs include the configuration model \citep{bender1978asymptotic, bollobas1979probabilistic, molloy1995critical}, the $\beta$-model \citep{chatterjee2011random}, and the Chung-Lu model \citep{aiello2000random}. As a special case, we consider the Chung-Lu fixed degree model with degree sequence $\mathbf{d}$. For $k = 1$, when $\theta_u = d(u) / \sqrt{\sum_{w \in [n]} d(w)}$, and $P_{1,1} = 1$, then the resulting expected edge weight between nodes $u, v \in [n]$ is given by:
	$$\mathbb{E}[w_{u,v}] = \dfrac{d(u) d(v)}{\sum_{w \in [n]} d(w)}.$$
This is precisely the expected edge weights associated with the Chung-Lu random graph model. The Chung-Lu model is often used as a null random graph model against which the features of an observed network is compared. For example, this model is often used for the detection and evaluation of community structure in networks \citep{newman2006modularity, wilson2013measuring, wilson2014testing}.
\end{itemize}

\subsection*{Appendix B: Additional Results for Simulation study}

We provide the average ARL for the simulations conducted in Section 7.2 for Phase I size $m = 25$ and $m = 50$ in Tables \ref{table:ARLm25} and \ref{table:ARLm50}, respectively.
\begin{table}
	\caption{\label{table:ARLm25} Average ARLs for Simulations in Section 7.2 when $m=25$.}
	\centering
	\begin{tabular}{cclllll}
		{\bf Simulation} & \multicolumn{2}{c}{\bf Change} & {$s$} & {${\widehat{P}_{1,1}}$} & {${\widehat{P}_{1,2}}$} & {${\widehat{P}_{2,2}}$} \\ \hline
		0 & & none & 425.50 & 507.53 & 512.40 & 534.40 \\ \hline
		  & & $\epsilon = 0.01$ & 474.20 & 299.50 & 487.64 & 506.20\\
		1 & $P^*_{1,1} = P_{1,1} + \epsilon$ & $\epsilon = 0.05$ & 613.67 & 19.75 & 494.40 & 482.20\\
		  & & $\epsilon = 0.10$ & 649.00 & 2.67  & 474.52 & 474.80\\ \hline
		  & & $\epsilon = 0.01$ & 587.50 & 280.60 & 149.50  & 297.00\\
		2 & $P^*_{i,j} = P_{i,j} + \epsilon$ & $\epsilon = 0.05$ & 555.70 & 18.44   & 1.98   & 17.84\\
		  & & $\epsilon = 0.10$ & 350.60  & 2.61   & 1.01   & 2.66\\ \hline		  
	      & &$\tau = 0.05$  & 366.70 & 383.30 & 482.40 & 490.17\\
	    3 & $\delta^*_1 = \delta_1 + \tau$ & $\tau = 0.10$  & 193.70 & 299.80 & 419.50 & 481.90\\
	      & & $\tau = 0.25$  & 35.80  & 118.90  & 306.00 & 509.40\\ \hline
  	      & & $\tau = 0.05$  & 229.90  & 380.00 & 480.70 & 382.80\\
  	    4 & $\delta^*_i = \delta_i + \tau$ & $\tau = 0.10$  & 107.90  & 312.90 & 342.00 & 288.50\\
  	      & &$\tau = 0.25$  & 6.98  & 132.90  & 188.50  & 108.00\\ \hline
    	  & &$n = 50$          & 451.90 & 288.30  & 1.90   & 271.30\\
    	5 & Merge comm. &$n = 100$         & 452.30 & 268.30  & 1.84   & 269.10\\
    	  & & $n = 500$         & 443.60  & 283.00  & 1.91   & 247.70\\ \hline
      	  & & $n = 50$          & 226.00 & 224.00  & 497.20 & 509.85\\
      	6 & Split comm. &$n = 100$         & 247.50 & 275.00  & 506.50 & 487.00\\
      	  & & $n = 500$         & 220.60  & 269.30  & 551.40 & 480.80\\ \hline
	\end{tabular}
\end{table}

\begin{table}
	\caption{\label{table:ARLm50} Average ARLs for Simulations in Section 7.2 when $m=50$.}
	\centering
	\begin{tabular}{cclllll}
		{\bf Simulation} & \multicolumn{2}{c}{\bf Change} & $s$ & {${\widehat{P}_{1,1}}$} & {${\widehat{P}_{1,2}}$} & {${\widehat{P}_{2,2}}$} \\ \hline
		0 & & none & 398.10 & 408.99 & 436.10 & 429.50 \\ \hline
		  & & $\epsilon = 0.01$ & 432.20 & 210.80 & 456.20 & 434.10\\
		1 & $P^*_{1,1} = P_{1,1} + \epsilon$ & $\epsilon = 0.05$ & 577.83 & 12.27   & 443.30 & 473.50\\
		  & & $\epsilon = 0.10$ & 604.00 & 2.36   & 435.99 & 424.50\\ \hline
		  & & $\epsilon = 0.01$ & 551.90 & 204.70 & 81.72  & 213.80\\
		2 & $P^*_{i,j} = P_{i,j} + \epsilon$ & $\epsilon = 0.05$ & 497.10 & 12.30   & 1.78   & 12.77\\
		  & & $\epsilon = 0.10$ & 217.10  & 2.34   & 1.01   & 2.34\\ \hline		  
	      & &$\tau = 0.05$  & 261.40 & 294.10 & 378.20 & 410.48\\
	    3 & $\delta^*_i = \delta_i + \tau$ & $\tau = 0.10$  & 136.68 & 225.10 & 359.90 & 427.50\\
	      & & $\tau = 0.25$  & 26.05  & 80.83  & 250.40 & 460.80\\ \hline
  	      & & $\tau = 0.05$  & 172.20  & 303.29 & 361.90 & 328.50\\
  	    4 & $\bm{\delta}^* = \bm{\delta} + \tau$ & $\tau = 0.10$  & 58.94  & 232.80 & 290.20 & 319.00\\
  	      & &$\tau = 0.25$  & 5.68   & 90.87  & 139.40  & 88.58\\ \hline
    	  & &$n = 50$          & 414.46 & 177.90  & 1.80   & 155.60\\
    	5 & Merge comm. &$n = 100$         & 366.99 & 171.90  & 1.88   & 169.50\\
    	  & & $n = 500$         & 386.70  & 142.90  & 1.83   & 162.60\\ \hline
      	  & & $n = 50$          & 172.10 & 165.10  & 472.40 & 457.50\\
      	6 & Split comm. &$n = 100$         & 163.90 & 145.00  & 480.90 & 436.02\\
      	  & & $n = 500$         & 169.50  & 165.30  & 428.15 & 424.00\\ \hline
	\end{tabular}
\end{table}

\end{document}